\documentclass[default]{aastex631}

\usepackage{comment}
\usepackage{longtable}
\usepackage{graphics}
\received{March 9, 2022}
\revised{May 8, 2022}
\accepted{May 10, 2022}
\submitjournal{ApJ}

\shorttitle{CALET Search for EM counterparts of GW during the LIGO/Virgo O3 run }
\shortauthors{Adriani et al.}
\graphicspath{{./}{figures/}}
\begin{document}

\title{CALET Search for electromagnetic counterparts of gravitational waves \\during the LIGO/Virgo O3 run \\}

\correspondingauthor{Yuta Kawakubo, Michael Cherry}
\email{kawakubo1@lsu.edu, cherry@lsu.edu}

%\author[0000-0002-0786-7307]{Yuta Kawakubo *** and other authors from ICRC list***}
%\affiliation{Dept. of Physics \& Astronomy, Louisiana State University, Baton Rouge, LA 70803, USA}
%\author{O. Adriani and other authors from the Nickel paper plus Val}
%\affiliation{from the list in the Nickel paper}

\author{O.~Adriani}
\affiliation{Department of Physics, University of Florence, Via Sansone, 1 - 50019, Sesto Fiorentino, Italy}
\affiliation{INFN Sezione di Florence, Via Sansone, 1 - 50019, Sesto Fiorentino, Italy}
\author{Y.~Akaike}
\affiliation{Waseda Research Institute for Science and Engineering, Waseda University, 17 Kikuicho,  Shinjuku, Tokyo 162-0044, Japan}
\affiliation{JEM Utilization Center, Human Spaceflight Technology Directorate, Japan Aerospace Exploration Agency, 2-1-1 Sengen, Tsukuba, Ibaraki 305-8505, Japan}
\author{K.~Asano}
\affiliation{Institute for Cosmic Ray Research, The University of Tokyo, 5-1-5 Kashiwa-no-Ha, Kashiwa, Chiba 277-8582, Japan}
\author{Y.~Asaoka}
\affiliation{Institute for Cosmic Ray Research, The University of Tokyo, 5-1-5 Kashiwa-no-Ha, Kashiwa, Chiba 277-8582, Japan}
\author{E.~Berti} 
\affiliation{Department of Physics, University of Florence, Via Sansone, 1 - 50019, Sesto Fiorentino, Italy}
\affiliation{INFN Sezione di Florence, Via Sansone, 1 - 50019, Sesto Fiorentino, Italy}
\author{G.~Bigongiari}
\affiliation{Department of Physical Sciences, Earth and Environment, University of Siena, via Roma 56, 53100 Siena, Italy}
\affiliation{INFN Sezione di Pisa, Polo Fibonacci, Largo B. Pontecorvo, 3 - 56127 Pisa, Italy}
\author{W.R.~Binns}
\affiliation{Department of Physics and McDonnell Center for the Space Sciences, Washington University, One Brookings Drive, St. Louis, Missouri 63130-4899, USA}
\author{M.~Bongi}
\affiliation{Department of Physics, University of Florence, Via Sansone, 1 - 50019, Sesto Fiorentino, Italy}
\affiliation{INFN Sezione di Florence, Via Sansone, 1 - 50019, Sesto Fiorentino, Italy}
\author{P.~Brogi}
\affiliation{Department of Physical Sciences, Earth and Environment, University of Siena, via Roma 56, 53100 Siena, Italy}
\affiliation{INFN Sezione di Pisa, Polo Fibonacci, Largo B. Pontecorvo, 3 - 56127 Pisa, Italy}
\author{A.~Bruno}
\affiliation{Heliospheric Physics Laboratory, NASA/GSFC, Greenbelt, Maryland 20771, USA}
\author{J.H.~Buckley}
\affiliation{Department of Physics and McDonnell Center for the Space Sciences, Washington University, One Brookings Drive, St. Louis, Missouri 63130-4899, USA}
\author{N.~Cannady}
\affiliation{Center for Space Sciences and Technology, University of Maryland, Baltimore County, 1000 Hilltop Circle, Baltimore, Maryland 21250, USA}
\affiliation{Astroparticle Physics Laboratory, NASA/GSFC, Greenbelt, Maryland 20771, USA}
\affiliation{Center for Research and Exploration in Space Sciences and Technology, NASA/GSFC, Greenbelt, Maryland 20771, USA}
\author{G.~Castellini}
\affiliation{Institute of Applied Physics (IFAC),  National Research Council (CNR), Via Madonna del Piano, 10, 50019, Sesto Fiorentino, Italy}
\author{C.~Checchia}
\affiliation{Department of Physical Sciences, Earth and Environment, University of Siena, via Roma 56, 53100 Siena, Italy}
\affiliation{INFN Sezione di Pisa, Polo Fibonacci, Largo B. Pontecorvo, 3 - 56127 Pisa, Italy}
\author{M.L.~Cherry}
\affiliation{Department of Physics and Astronomy, Louisiana State University, 202 Nicholson Hall, Baton Rouge, Louisiana 70803, USA}
\author{G.~Collazuol}
\affiliation{Department of Physics and Astronomy, University of Padova, Via Marzolo, 8, 35131 Padova, Italy}
\affiliation{INFN Sezione di Padova, Via Marzolo, 8, 35131 Padova, Italy} 
\author{K.~Ebisawa}
\affiliation{Institute of Space and Astronautical Science, Japan Aerospace Exploration Agency, 3-1-1 Yoshinodai, Chuo, Sagamihara, Kanagawa 252-5210, Japan}
\author{A.~W.~Ficklin}
\affiliation{Department of Physics and Astronomy, Louisiana State University, 202 Nicholson Hall, Baton Rouge, Louisiana 70803, USA}
\author{H.~Fuke}
\affiliation{Institute of Space and Astronautical Science, Japan Aerospace Exploration Agency, 3-1-1 Yoshinodai, Chuo, Sagamihara, Kanagawa 252-5210, Japan}
\author{S.~Gonzi}
\affiliation{Department of Physics, University of Florence, Via Sansone, 1 - 50019, Sesto Fiorentino, Italy}
\affiliation{INFN Sezione di Florence, Via Sansone, 1 - 50019, Sesto Fiorentino, Italy}
\author{T.G.~Guzik}
\affiliation{Department of Physics and Astronomy, Louisiana State University, 202 Nicholson Hall, Baton Rouge, Louisiana 70803, USA}
\author{T.~Hams}
\affiliation{Center for Space Sciences and Technology, University of Maryland, Baltimore County, 1000 Hilltop Circle, Baltimore, Maryland 21250, USA}
\author{K.~Hibino}
\affiliation{Kanagawa University, 3-27-1 Rokkakubashi, Kanagawa, Yokohama, Kanagawa 221-8686, Japan}
\author{M.~Ichimura}
\affiliation{Faculty of Science and Technology, Graduate School of Science and Technology, Hirosaki University, 3, Bunkyo, Hirosaki, Aomori 036-8561, Japan}
\author{K.~Ioka}
\affiliation{Yukawa Institute for Theoretical Physics, Kyoto University, Kitashirakawa Oiwake-cho, Sakyo-ku, Kyoto, 606-8502, Japan}
\author{W.~Ishizaki}
\affiliation{Institute for Cosmic Ray Research, The University of Tokyo, 5-1-5 Kashiwa-no-Ha, Kashiwa, Chiba 277-8582, Japan}
\author{M.H.~Israel}
\affiliation{Department of Physics and McDonnell Center for the Space Sciences, Washington University, One Brookings Drive, St. Louis, Missouri 63130-4899, USA}
\author{K.~Kasahara}
\affiliation{Department of Electronic Information Systems, Shibaura Institute of Technology, 307 Fukasaku, Minuma, Saitama 337-8570, Japan}
\author{J.~Kataoka}
\affiliation{School of Advanced Science and Engineering, Waseda University, 3-4-1 Okubo, Shinjuku, Tokyo 169-8555, Japan}
\author{R.~Kataoka}
\affiliation{National Institute of Polar Research, 10-3, Midori-cho, Tachikawa, Tokyo 190-8518, Japan}
\author{Y.~Katayose}
\affiliation{Faculty of Engineering, Division of Intelligent Systems Engineering, Yokohama National University, 79-5 Tokiwadai, Hodogaya, Yokohama 240-8501, Japan}
\author{C.~Kato}
\affiliation{Faculty of Science, Shinshu University, 3-1-1 Asahi, Matsumoto, Nagano 390-8621, Japan}
\author{N.~Kawanaka}
\affiliation{Yukawa Institute for Theoretical Physics, Kyoto University, Kitashirakawa Oiwake-cho, Sakyo-ku, Kyoto, 606-8502, Japan}
\author{Y.~Kawakubo}
\affiliation{Department of Physics and Astronomy, Louisiana State University, 202 Nicholson Hall, Baton Rouge, Louisiana 70803, USA}
\author{K.~Kobayashi}
\affiliation{Waseda Research Institute for Science and Engineering, Waseda University, 17 Kikuicho,  Shinjuku, Tokyo 162-0044, Japan}
\affiliation{JEM Utilization Center, Human Spaceflight Technology Directorate, Japan Aerospace Exploration Agency, 2-1-1 Sengen, Tsukuba, Ibaraki 305-8505, Japan}
\author{K.~Kohri} 
\affiliation{Institute of Particle and Nuclear Studies, High Energy Accelerator Research Organization, 1-1 Oho, Tsukuba, Ibaraki, 305-0801, Japan} 
\author{H.S.~Krawczynski}
\affiliation{Department of Physics and McDonnell Center for the Space Sciences, Washington University, One Brookings Drive, St. Louis, Missouri 63130-4899, USA}
\author{J.F.~Krizmanic}
\affiliation{Astroparticle Physics Laboratory, NASA/GSFC, Greenbelt, Maryland 20771, USA}
\author{P.~Maestro}
\affiliation{Department of Physical Sciences, Earth and Environment, University of Siena, via Roma 56, 53100 Siena, Italy}
\affiliation{INFN Sezione di Pisa, Polo Fibonacci, Largo B. Pontecorvo, 3 - 56127 Pisa, Italy}
\author{P.S.~Marrocchesi}
\affiliation{Department of Physical Sciences, Earth and Environment, University of Siena, via Roma 56, 53100 Siena, Italy}
\affiliation{INFN Sezione di Pisa, Polo Fibonacci, Largo B. Pontecorvo, 3 - 56127 Pisa, Italy}
\author{A.M.~Messineo}
\affiliation{University of Pisa, Polo Fibonacci, Largo B. Pontecorvo, 3 - 56127 Pisa, Italy}
\affiliation{INFN Sezione di Pisa, Polo Fibonacci, Largo B. Pontecorvo, 3 - 56127 Pisa, Italy}
\author{J.W.~Mitchell}
\affiliation{Astroparticle Physics Laboratory, NASA/GSFC, Greenbelt, Maryland 20771, USA}
\author{S.~Miyake}
\affiliation{Department of Electrical and Electronic Systems Engineering, National Institute of Technology (KOSEN), Ibaraki College, 866 Nakane, Hitachinaka, Ibaraki 312-8508, Japan}
\author{A.A.~Moiseev}
\affiliation{Department of Astronomy, University of Maryland, College Park, Maryland 20742, USA}
\affiliation{Astroparticle Physics Laboratory, NASA/GSFC, Greenbelt, Maryland 20771, USA}
\affiliation{Center for Research and Exploration in Space Sciences and Technology, NASA/GSFC, Greenbelt, Maryland 20771, USA}
\author{M.~Mori}
\affiliation{Department of Physical Sciences, College of Science and Engineering, Ritsumeikan University, Shiga 525-8577, Japan}
\author{N.~Mori}
\affiliation{INFN Sezione di Florence, Via Sansone, 1 - 50019, Sesto Fiorentino, Italy}
\author{H.M.~Motz}
\affiliation{Faculty of Science and Engineering, Global Center for Science and Engineering, Waseda University, 3-4-1 Okubo, Shinjuku, Tokyo 169-8555, Japan}
\author{K.~Munakata}
\affiliation{Faculty of Science, Shinshu University, 3-1-1 Asahi, Matsumoto, Nagano 390-8621, Japan}
\author{S.~Nakahira}
\affiliation{Institute of Space and Astronautical Science, Japan Aerospace Exploration Agency, 3-1-1 Yoshinodai, Chuo, Sagamihara, Kanagawa 252-5210, Japan}
\author{J.~Nishimura}
\affiliation{Institute of Space and Astronautical Science, Japan Aerospace Exploration Agency, 3-1-1 Yoshinodai, Chuo, Sagamihara, Kanagawa 252-5210, Japan}
\author{G.A.~de~Nolfo}
\affiliation{Heliospheric Physics Laboratory, NASA/GSFC, Greenbelt, Maryland 20771, USA}
\author{S.~Okuno}
\affiliation{Kanagawa University, 3-27-1 Rokkakubashi, Kanagawa, Yokohama, Kanagawa 221-8686, Japan}
\author{J.F.~Ormes}
\affiliation{Department of Physics and Astronomy, University of Denver, Physics Building, Room 211, 2112 East Wesley Avenue, Denver, Colorado 80208-6900, USA}
\author{N.~Ospina}
\affiliation{Department of Physics and Astronomy, University of Padova, Via Marzolo, 8, 35131 Padova, Italy}\affiliation{INFN Sezione di Padova, Via Marzolo, 8, 35131 Padova, Italy} 
\author{S.~Ozawa}
\affiliation{Quantum ICT Advanced Development Center, National Institute of Information and Communications Technology, 4-2-1 Nukui-Kitamachi, Koganei, Tokyo 184-8795, Japan}
\author{L.~Pacini}
\affiliation{Department of Physics, University of Florence, Via Sansone, 1 - 50019, Sesto Fiorentino, Italy}
\affiliation{Institute of Applied Physics (IFAC),  National Research Council (CNR), Via Madonna del Piano, 10, 50019, Sesto Fiorentino, Italy}
\affiliation{INFN Sezione di Florence, Via Sansone, 1 - 50019, Sesto Fiorentino, Italy}
\author{P.~Papini}
\affiliation{INFN Sezione di Florence, Via Sansone, 1 - 50019, Sesto Fiorentino, Italy}
\author{B.F.~Rauch}
\affiliation{Department of Physics and McDonnell Center for the Space Sciences, Washington University, One Brookings Drive, St. Louis, Missouri 63130-4899, USA}
\author{S.B.~Ricciarini}
\affiliation{Institute of Applied Physics (IFAC),  National Research Council (CNR), Via Madonna del Piano, 10, 50019, Sesto Fiorentino, Italy}
\affiliation{INFN Sezione di Florence, Via Sansone, 1 - 50019, Sesto Fiorentino, Italy}
\author{K.~Sakai}
\affiliation{Center for Space Sciences and Technology, University of Maryland, Baltimore County, 1000 Hilltop Circle, Baltimore, Maryland 21250, USA}
\affiliation{Astroparticle Physics Laboratory, NASA/GSFC, Greenbelt, Maryland 20771, USA}
\affiliation{Center for Research and Exploration in Space Sciences and Technology, NASA/GSFC, Greenbelt, Maryland 20771, USA}
\author{T.~Sakamoto}
\affiliation{College of Science and Engineering, Department of Physics and Mathematics, Aoyama Gakuin University,  5-10-1 Fuchinobe, Chuo, Sagamihara, Kanagawa 252-5258, Japan}
\author{M.~Sasaki}
\affiliation{Department of Astronomy, University of Maryland, College Park, Maryland 20742, USA}
\affiliation{Astroparticle Physics Laboratory, NASA/GSFC, Greenbelt, Maryland 20771, USA}
\affiliation{Center for Research and Exploration in Space Sciences and Technology, NASA/GSFC, Greenbelt, Maryland 20771, USA}
\author{Y.~Shimizu}
\affiliation{Kanagawa University, 3-27-1 Rokkakubashi, Kanagawa, Yokohama, Kanagawa 221-8686, Japan}
\author{A.~Shiomi}
\affiliation{College of Industrial Technology, Nihon University, 1-2-1 Izumi, Narashino, Chiba 275-8575, Japan}
\author{P.~Spillantini}
\affiliation{Department of Physics, University of Florence, Via Sansone, 1 - 50019, Sesto Fiorentino, Italy}
\author{F.~Stolzi}
\affiliation{Department of Physical Sciences, Earth and Environment, University of Siena, via Roma 56, 53100 Siena, Italy}
\affiliation{INFN Sezione di Pisa, Polo Fibonacci, Largo B. Pontecorvo, 3 - 56127 Pisa, Italy}
\author{S.~Sugita}
\affiliation{College of Science and Engineering, Department of Physics and Mathematics, Aoyama Gakuin University,  5-10-1 Fuchinobe, Chuo, Sagamihara, Kanagawa 252-5258, Japan}
\author{A.~Sulaj} 
\affiliation{Department of Physical Sciences, Earth and Environment, University of Siena, via Roma 56, 53100 Siena, Italy}
\affiliation{INFN Sezione di Pisa, Polo Fibonacci, Largo B. Pontecorvo, 3 - 56127 Pisa, Italy}
\author{M.~Takita}
\affiliation{Institute for Cosmic Ray Research, The University of Tokyo, 5-1-5 Kashiwa-no-Ha, Kashiwa, Chiba 277-8582, Japan}
\author{T.~Tamura}
\affiliation{Kanagawa University, 3-27-1 Rokkakubashi, Kanagawa, Yokohama, Kanagawa 221-8686, Japan}
\author{T.~Terasawa}
\affiliation{Institute for Cosmic Ray Research, The University of Tokyo, 5-1-5 Kashiwa-no-Ha, Kashiwa, Chiba 277-8582, Japan}
\author{S.~Torii}
\affiliation{Waseda Research Institute for Science and Engineering, Waseda University, 17 Kikuicho,  Shinjuku, Tokyo 162-0044, Japan}
\author{Y.~Tsunesada}
%\affiliation{Division of Mathematics and Physics, Graduate School of Science, Osaka City University, 3-3-138 Sugimoto, Sumiyoshi, Osaka 558-8585, Japan}
\affiliation{Graduate School of Science, Osaka Metropolitan University, Sugimoto, Sumiyoshi, Osaka 558-8585, Japan}
%\affiliation{Nambu Yoichiro Institute of Theoretical and Experimental Physics, Osaka City University, 3-3-138 Sugimoto, Sumiyoshi, Osaka 558-8585, Japan }
\affiliation{Nambu Yoichiro Institute for Theoretical and Experimental Physics, Osaka Metropolitan University,  Sugimoto, Sumiyoshi, Osaka 558-8585, Japan}

\author{Y.~Uchihori}
\affiliation{National Institutes for Quantum and Radiation Science and Technology, 4-9-1 Anagawa, Inage, Chiba 263-8555, Japan}
\author{E.~Vannuccini}
\affiliation{INFN Sezione di Florence, Via Sansone, 1 - 50019, Sesto Fiorentino, Italy}
\author{J.P.~Wefel}
\affiliation{Department of Physics and Astronomy, Louisiana State University, 202 Nicholson Hall, Baton Rouge, Louisiana 70803, USA}
\author{K.~Yamaoka}
\affiliation{Nagoya University, Furo, Chikusa, Nagoya 464-8601, Japan}
\author{S.~Yanagita}
\affiliation{College of Science, Ibaraki University, 2-1-1 Bunkyo, Mito, Ibaraki 310-8512, Japan}
\author{A.~Yoshida}
\affiliation{College of Science and Engineering, Department of Physics and Mathematics, Aoyama Gakuin University,  5-10-1 Fuchinobe, Chuo, Sagamihara, Kanagawa 252-5258, Japan}
\author{K.~Yoshida}
\affiliation{Department of Electronic Information Systems, Shibaura Institute of Technology, 307 Fukasaku, Minuma, Saitama 337-8570, Japan}
\author{W.~V.~Zober}
\affiliation{Department of Physics and McDonnell Center for the Space Sciences, Washington University, One Brookings Drive, St. Louis, Missouri 63130-4899, USA}

%\collaboration{1}{(CALET collaboration)}
%\author{Butler Burton}
%\affiliation{Leiden University}
%\affiliation{AAS Journals Associate Editor-in-Chief}
%\nocollaboration{1}



\begin{abstract}

The CALorimetric Electron Telescope (CALET) on the International Space Station (ISS) consists of a high-energy cosmic ray CALorimeter (CAL) and a lower-energy  CALET Gamma ray Burst Monitor (CGBM). CAL is sensitive to electrons up to 20 TeV, cosmic ray nuclei from Z = 1 through Z $\sim 40$, and gamma rays over the range 1 GeV - 10 TeV. CGBM observes  gamma rays from 7 keV to 20 MeV. The combined CAL-CGBM instrument has conducted a search for gamma ray bursts (GRBs) since Oct. 2015. We report here on the results of a search for X-ray/gamma ray counterparts to gravitational wave events reported during the LIGO/Virgo observing run O3. No events have been detected that pass all acceptance criteria. We describe the components, performance, and triggering algorithms of the CGBM -- the two Hard X-ray Monitors (HXM) consisting of LaBr$_3$(Ce) scintillators sensitive to 7 keV to 1 MeV gamma rays and a Soft Gamma ray Monitor (SGM) BGO scintillator sensitive to 40 keV to 20 MeV -- and the high-energy CAL consisting of a CHarge-Detection module (CHD), IMaging Calorimeter (IMC), and fully active Total Absorption Calorimeter (TASC).  The analysis procedure is described and upper limits to the time-averaged fluxes are presented.

%This example manuscript is intended to serve as a tutorial and template for
%authors to use when writing their own AAS Journal articles. The manuscript
%includes a history of \aastex\ and documents the new features in the
%previous versions as well as the new features in version 6.3. This
%manuscript includes many figure and table examples to illustrate these new
%features.  Information on features not explicitly mentioned in the article
%can be viewed in the manuscript comments or more extensive online
%documentation. Authors are welcome replace the text, tables, figures, and
%bibliography with their own and submit the resulting manuscript to the AAS
%Journals peer review system.  The first lesson in the tutorial is to remind
%authors that the AAS Journals, the Astrophysical Journal (ApJ), the
%Astrophysical Journal Letters (ApJL), and Astronomical Journal (AJ), all
%have a 250 word limit for the abstract\footnote{Note that manuscripts 
%submitted to the new Research Notes of the American Astronomical Society 
%(RNAAS) do \textbf{not} have abstracts.}.  If you exceed this length the
%Editorial office will ask you to shorten it. This abstract has 180 words.

\end{abstract}

%\KeyWords{gravitational waves \-\-\- gamma rays \-\-\-  other key words}

\section{Introduction}
The importance of simultaneous or near-simultaneous multi-messenger observations has long been recognized (e.g. \cite{Meszaros}  and \cite{Burns_Multimessenger} for recent reviews), and was clearly demonstrated over three decades ago by the correlated neutrino and optical observations of SN1987A \citep{Kamiokande,IMB,SN1987A,Baksan}. The observations of the gravitational radiation event GW 170817 from a binary  neutron star merger \citep{LIGO-170817} together with the short gamma ray burst GRB 170817A \citep{GBM-170817, INTEGRAL-170817} and the optical transient AT2017gfo  \citep{joint_observation-170817} have now made it possible to draw physics conclusions about the connection between short GRBs and neutron star mergers, the origin of heavy r-process elements, the speed of gravitational waves, and the nature of kilonovae. Additional observations of short GRBs associated
with gravitational wave events will especially provide information about neutron stars and their mergers, the nature of the GRB jet, and the neutron star equation of state \citep{Burns_neutron_stars,Pian}. A number of joint gamma ray/gravitational wave searches have been carried out since the 2017 event involving LIGO/Virgo and {\it Fermi}-GBM \citep{GBM-LIGO-O1and2,LIGO_Single_events}, {\it Swift} \citep{Swift,LIGO-GBM-Swift}, {\it INTEGRAL} \citep{INTEGRAL},  {\it AGILE} \citep{AGILE}, and CALET \citep{Yamaoka-2017,CALET-O1and2}. In no case since GRB 170817A/GW 170817, however, has there been a confirmed positive GRB signal in association with a LIGO/Virgo gravitational wave event.

\begin{figure}[htbp]
\begin{center}
\includegraphics[width=0.7\textwidth]{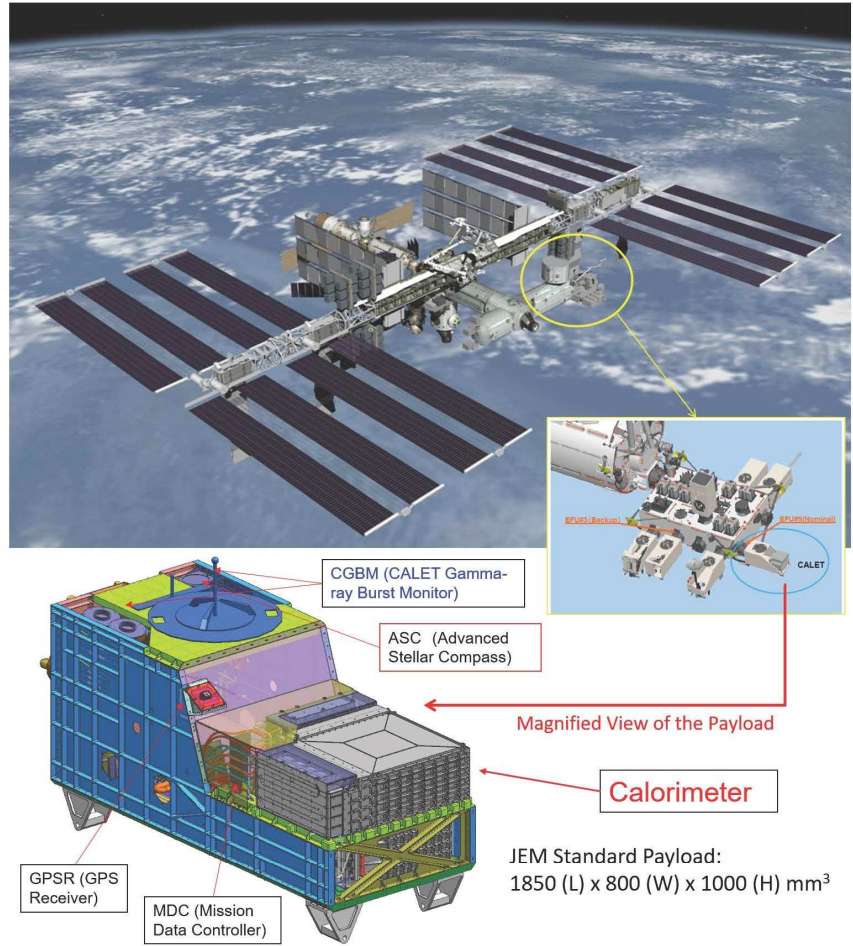}
\caption{\label{fig:CALET_on_ISS} Schematic of CALET on the ISS. CALET is mounted on port 9 of JEM-EF.  CALET consists of CAL, CGBM, support sensors including the Advanced Stellar Camera (ASC) and the Global Position Sensor Receiver (GPSR), and the Mission Data Controller (MDC) which controls the CALET detectors and acquires the data from the instruments.}
\end{center}
\end{figure}

The Japanese-Italian-US CALET cosmic ray/gamma ray telescope (Fig. \ref{fig:CALET_on_ISS})
was launched to the International Space Station (ISS) on August 19, 2015 and has been in operation on the Exposed Facility of the Japanese Experiment Module (JEM-EF) of the ISS since October 2015 \citep{CALET-electrons, CALET-protons,  CALET-CO, CALET-iron}. 
The main detector of the CALET payload is a calorimeter (CAL) to observe high-energy cosmic rays and gamma rays above 1 GeV. In addition, the Gamma ray Burst monitor (CGBM)  covers the gamma ray energy range from 7 keV to 20 MeV. 

The ongoing CGBM and CAL searches for GRBs are described by \cite{CALET-GRBs-2021}. The searches for CGBM and CAL counterparts to Advanced LIGO and Advanced Virgo gravitational wave events during LIGO/Virgo observing runs O1 and O2 are described in \cite{Yamaoka-2017} and  \cite{CALET-O1and2}.
The present paper describes the search by the CALET gamma ray detectors for prompt GRBs associated with gravitational wave events during LIGO/Virgo observing run O3. In Sec. 2, we describe the low-energy and high-energy CALET gamma ray telescopes; in Sec. 3 we discuss the analysis procedures and present the results of the CALET GRB counterpart search during LIGO/Virgo observation run O3.  Finally, we discuss results and conclusions in Sec. 4.

\section{CALET and gamma ray burst observations}

\subsection{CALET Gamma ray Burst Monitor (CGBM)}

\begin{figure}[htbp]
\begin{center}
\includegraphics[width=0.7\textwidth]{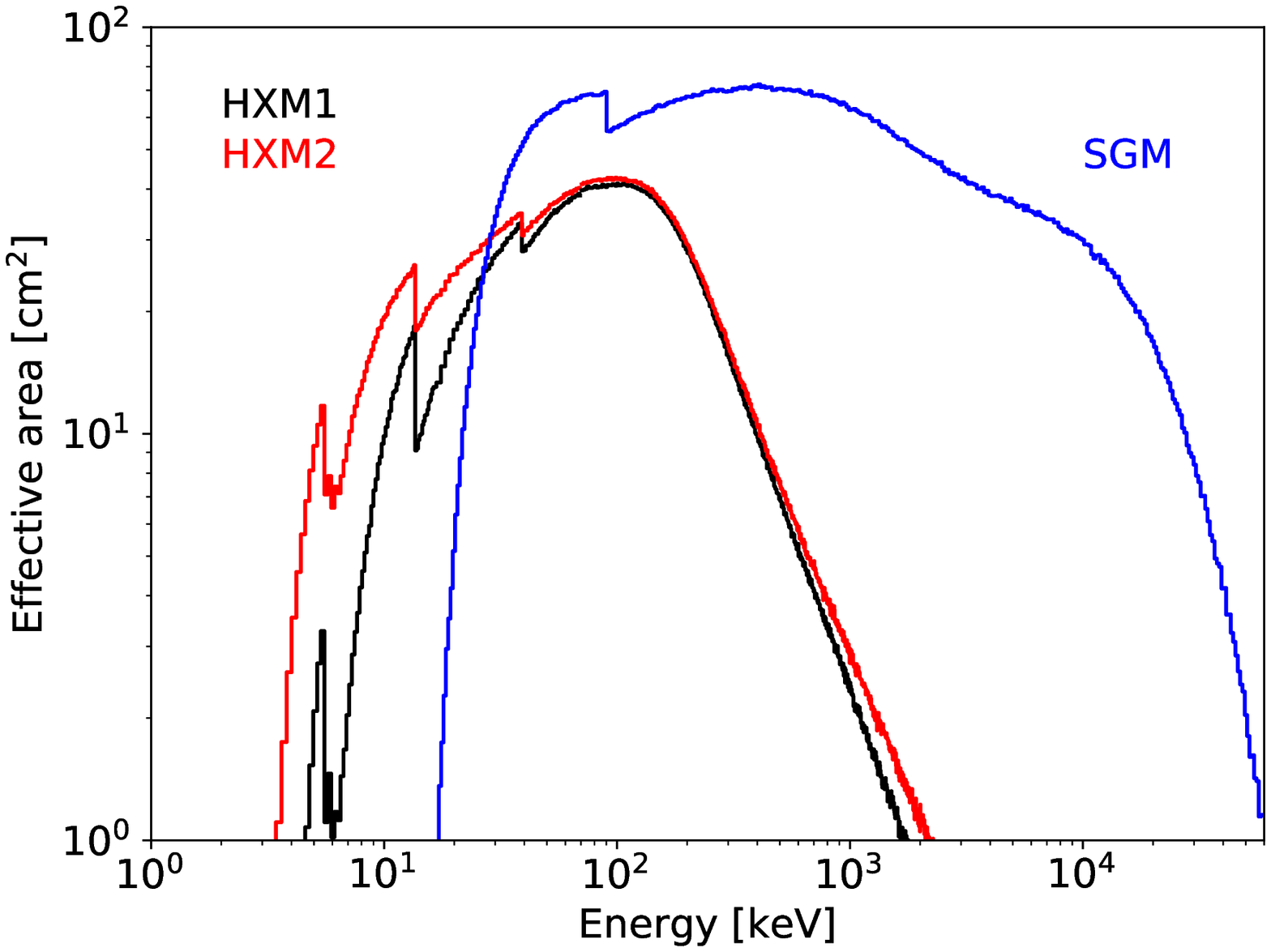}
\caption{\label{fig:CGBM_effective_area} CGBM effective areas vs gamma ray energy for the individual HXM1 and HXM2 detectors and for SGM at vertical incidence.}	
\end{center}
\end{figure}

The CAL and CGBM instruments have gamma ray sensitivity in different energy ranges.  CGBM is primarily responsible for observing prompt emission from GRBs \citep{CALET-GRBs-2019, CALET-GRBs-2021}.  CGBM consists of two Hard X-ray Monitors (HXMs) and the Soft Gamma ray Monitor (SGM). Each HXM module consists of a  6.6 cm diameter $\times$ 1.3 cm deep and a 7.9 cm diameter $\times$ 1.3 cm deep LaBr$_3$(Ce) scintillator; SGM consists of a 10.2 cm diameter $\times$ 7.6 cm deep  BGO scintillator. Both are viewed by photomultiplier tubes.  Outputs of the photomultiplier tubes are amplified by two amplifiers with a ratio of gains $\sim 30$ and pulse heights are acquired individually as High Gain and Low Gain Pulse Heights in order to provide the required dynamic range.  The two subsystems  provide sensitivity to X-rays and gamma rays over the energy range  7 keV - 1 MeV (HXM) and 40 keV - 20 MeV (SGM), as  shown in Fig. \ref{fig:CGBM_effective_area}.  The detectors have fields of view $\sim 3$ sr (HXM) and $\sim 8$ sr (SGM). 

CGBM normally collects two types of continuous monitor data suitable for temporal analysis and spectral analysis:  time history (TH) data with 0.125 s resolution and 8 energy channels (4 channels in High Gain and 4 channels in Low Gain) and pulse height (PH) data with  4 s resolution and 512 channels (102 channels with High Gain and 410 channels with Low Gain).  Energy ranges for the High Gain and Low Gain channels in TH data are listed in Table \ref{tab:TH_energy}.  Monitor data are collected every 0.125 s and transmitted to the ground every second. In addition, an on-board trigger system  uses the signal-to-noise ratio (SNR) calculated every 0.25 s to detect increased count rates:

\begin{equation}
\label{eq:GRB_logic}
\mathrm{SNR} = \frac{N_{\mathrm{tot}} - \frac{N_{\mathrm{BG}}}{\Delta t_{\mathrm{BG}}}\Delta t}{\sqrt{\frac{N_{\mathrm{BG}}}{\Delta t_{\mathrm{BG}}}\Delta t } } \; \; \; .
\end{equation}

\noindent Here $\Delta t$ is the foreground (signal) integration time;  $N_{\mathrm{tot}}$ is the number of counts integrated over $\Delta t$ in the selected energy range; and $N_{\mathrm{BG}}$ is the number of background counts in the selected energy range integrated over the background time interval  $\Delta t_{\mathrm{BG}}$ preceding $\Delta t$.

SNRs are calculated on-board continuously every 0.25 s for  $\Delta t_{\mathrm{BG}}$ = 16 s  and four signal integration times ($\Delta t$ = 0.25 s, 0.5 s, 1 s, or 4 s) over the energy ranges 25 - 100 keV for HXM and 50 - 300 keV for SGM. An on-board trigger threshold is set at SNR = 8.5 for each HXM and 7.0 for SGM. 
If any calculated SNR exceeds the trigger thresholds, CGBM captures $\sim 700$ s of  individual event data starting $\sim 200$ s prior to the trigger and consisting of event arrival times with time resolution of 62.5 $\mu$s and ADC pulse heights corresponding to the individual energy deposits in each detector. Each event consists of two ADC pulse heights measured by High Gain and Low Gain.  When a CGBM on-board trigger occurs, 1) the CGBM event-by-event data with fine time resolution are recorded; 2) the energy threshold of the  CAL is reduced from its normal 10 GeV threshold to 1 GeV for approximately 10 minutes (Low Energy Gamma ray mode) to increase the  CAL sensitivity to GeV photons from GRBs; and 3) two optical images with an exposure of 0.5 sec are taken by the Advanced Stellar Compass (ASC) star sensor \citep{ASC} to detect optical flashes during the prompt emission phase. The CGBM event data buffer can store event data from four successive triggers on board. When a fourth trigger occurs before the buffer is emptied, the on-board trigger is disabled until event data are downlinked and deleted.  The downlinks of the CGBM event data and optical images are performed three times a week.

%\newpage
\begin{table*}[t]
\caption{Energy ranges of TH channels}
\label{tab:TH_energy}
\begin{center}
\small
\begin{tabular}{lcc} \hline\hline\\[-6pt]
TH channel & HXM & SGM\\  \hline
High gain ch0 & 7 - 10 keV & 40 - 100 keV\\
High gain ch1 & 10 - 25 keV & 100 - 230 keV\\
High gain ch2 & 25 - 50 keV & 230 - 450 keV\\
High gain ch3 & 50 - 100 keV & 450 - 1000 keV\\
Low gain ch0 & 60 - 100 keV & 550 - 830 keV \\
Low gain ch1 & 100 - 170 keV & 830 - 1500 keV\\
Low gain ch2 & 170 - 300 keV & 1.5 - 2.6 MeV\\
Low gain ch3 & 300 - 3000 keV & 2.6 - 28 MeV\\
\hline 
\end{tabular}
\end{center}
\end{table*}

A trigger alert system running on the ground server at the Tsukuba Space Center of the Japanese Aerospace Exploration Agency (JAXA) analyzes the  real-time monitoring data including housekeeping data, status information, and data settings when a CGBM on-board trigger occurs \citep{ground_ops}. When the alert system notices a CGBM on-board trigger, the real-time TH data are analyzed and a GCN notice is delivered.  If real time data are unavailable due to loss of connection between the  ISS and ground, hourly data distributed with at most one hour delay can be used for ground analysis.

\begin{figure}[htbp]
\begin{center}
\includegraphics[width=0.8\textwidth]{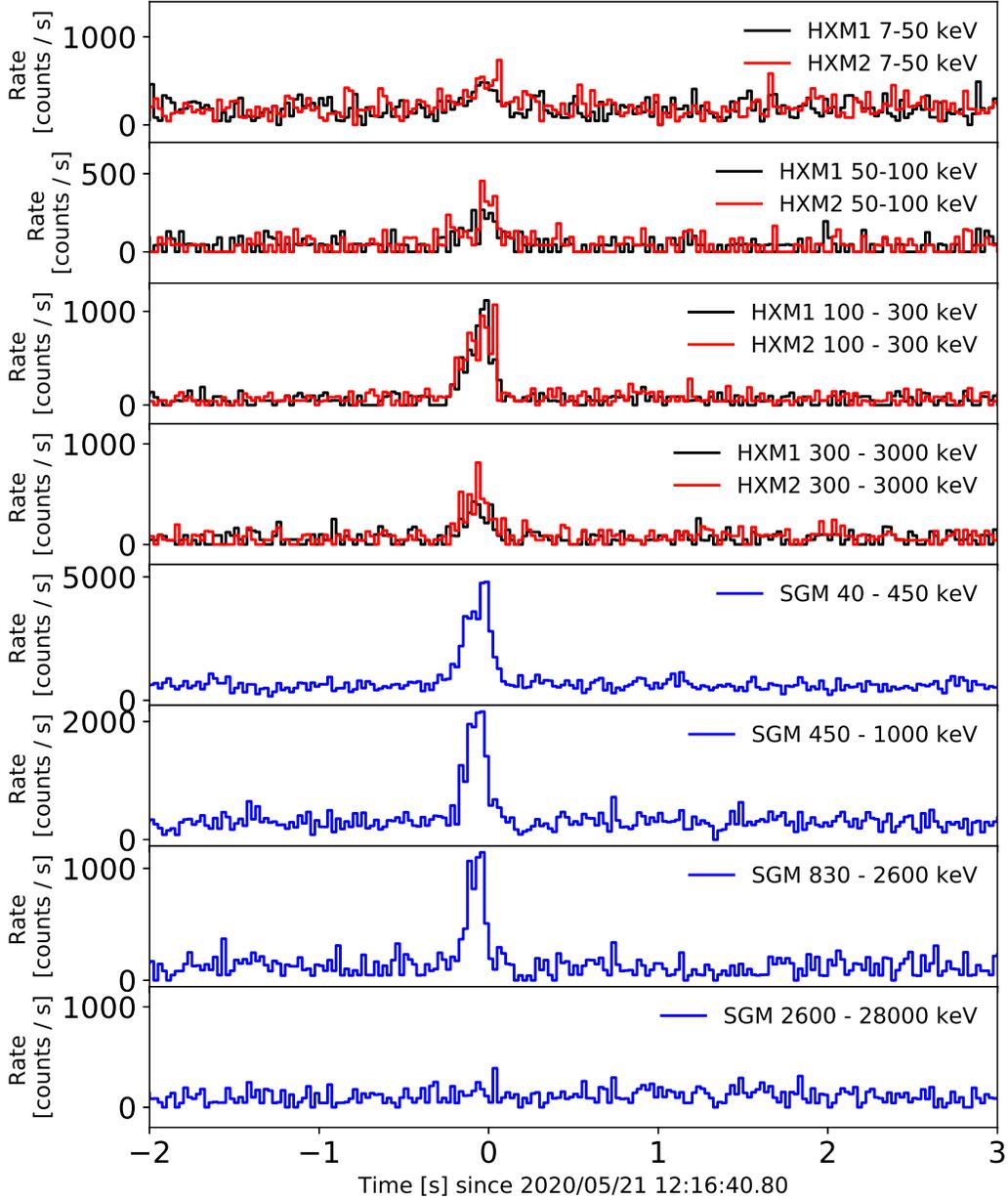}
\caption{\label{fig:GRB 200521A} Time histories of counts observed in HXM1, HXM2, and SGM for GRB 200521A.}
\end{center}
\end{figure}

\begin{figure}[htbp]
\begin{center}
\includegraphics[width=0.5\textwidth]{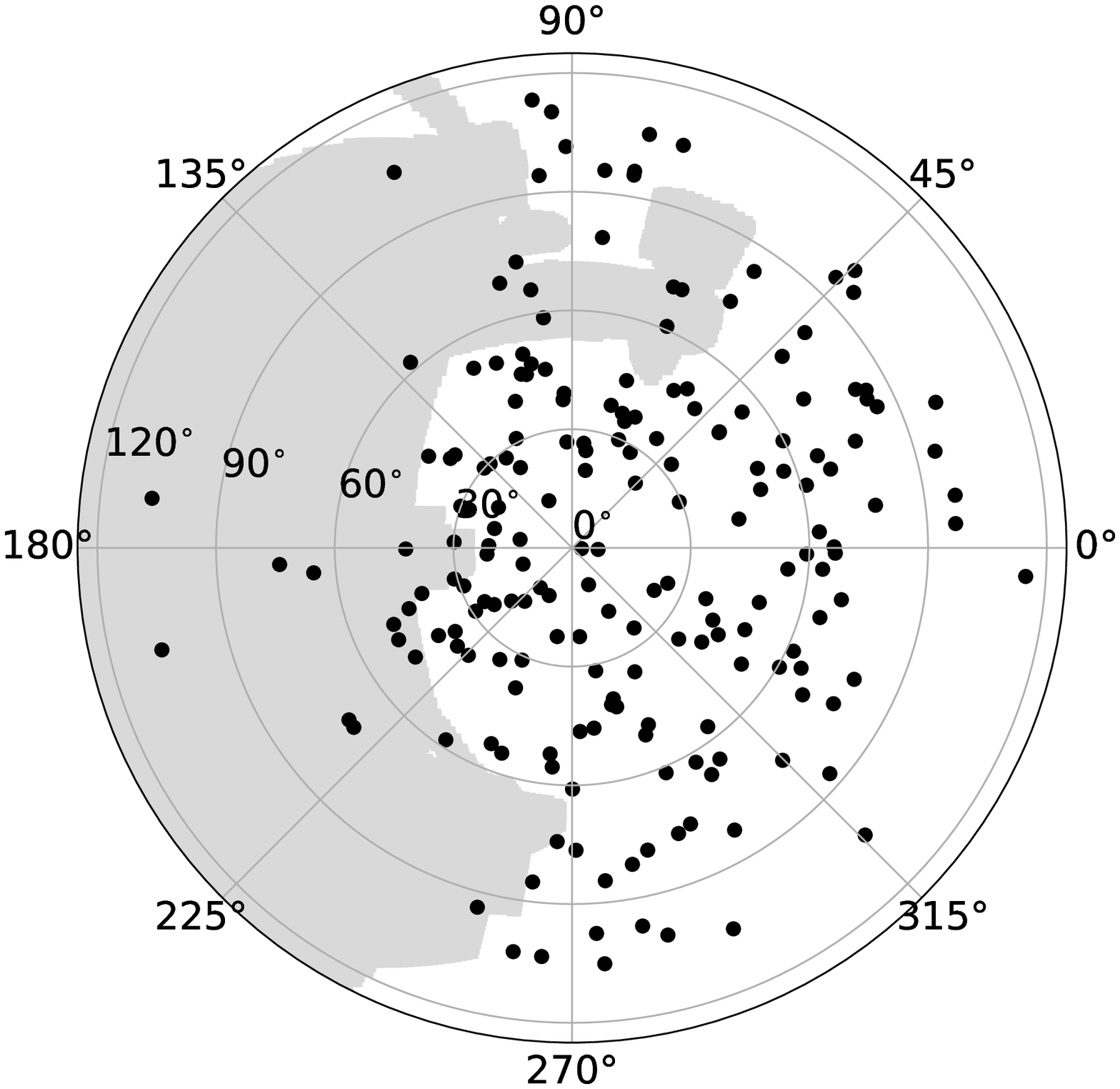}
\caption{\label{CGBM_sky} Incident angle distribution of
GRBs in the SGM field of view. Black points are GRB positions in the SGM coordinate. Gray shaded regions show the ISS fixed structures viewed from CALET.}
\end{center}
\end{figure}

As an example, time histories of GRB 200521A observed in the three detectors of CGBM are shown in Fig. \ref{fig:GRB 200521A} \citep{GRB200521A_CALET}. GRB 200521A was also detected by ASIM, {\it AGILE}, {\it INTEGRAL} SPI-ACS, and Konus-{\it Wind}  \citep{GRB200521A_ASIM, GRB200521A_AGILE, GRB200521A_KONUS}. A sky map of GRBs detected by CGBM through November 2021 is shown in Fig. \ref{CGBM_sky}.
Since CGBM has no capability to localize the GRBs,  GRB positions were based on the reports to the Gamma ray Coordinates Network (GCN, https://gcn.gsfc.nasa.gov) by other GRB instruments. Out of  271 GRBs detected by CGBM, 195 were localized by other  instruments.

Charged particles trapped by the Earth’s geomagnetic field cause the CGBM count rates to increase at  high geomagnetic latitudes and through the South Atlantic Anomaly (SAA). Therefore, CGBM high voltages are turned off at  high geomagnetic latitudes and during SAA passages in order to avoid false triggers  and excessive PMT currents due to increased particle fluxes. As a result, the  cumulative CGBM live time over the period Oct. 2015 - November 2021 is $\sim 60\%$.

The Active Thermal Control System (ATCS) of the JEM-EF maintains temperature variations for CAL to within a few degrees. However, CGBM temperatures are not controlled by the ATCS. CGBM temperature variations depend on  both the solar beta angle and solar altitude. Temperature corrections can be made on the ground using prominent background lines at 34 keV, 511 keV, 1.47 MeV, and 2.2 MeV.  

\subsection{CALET Calorimeter (CAL)}
The primary purpose of CAL is to observe high-energy electrons, protons and nuclei. In addition, CAL is also sensitive to gamma rays at 1~GeV - 10~TeV \citep{CALET-gammas}.  The primary CAL trigger mode for cosmic rays and gamma rays is the High-Energy (HE) mode with a threshold of 10 GeV for the observed energy.  CAL  typically takes data in HE mode, but when the geomagnetic latitude is below 20$^{\circ}$ (except around the SAA) or following a CGBM trigger, CAL adds Low-Energy Gamma-ray (LEG) mode with the gamma-ray trigger threshold  lowered to 1 GeV \citep{CALET-gammas}.

CAL identifies particles based on the ionization energy deposited in three separate detector systems: 1) The CHarge Detector(CHD)  located at the top of CAL consists of two orthogonal layers made of 14 plastic scintillator paddles with individual element charge resolution for particles from electrons and protons to ultra-heavy nuclei with  $Z \sim 40$.  2) Below the CHD, a finely segmented preshower IMaging Calorimeter (IMC) consists of eight double layers of 1 mm$^2$ cross section scintillating fibers, arranged in belts along orthogonal directions, interspersed with seven layers of tungsten with a total thickness of 3 radiation lengths (r.l.).  IMC can observe tracks of incident particles and provides an independent charge determination, fine-grained tracking information, and an image of the  initial stage of the shower development.  3) The Total Absorption Calorimeter (TASC) located at the bottom of CAL consists of  lead tungstate (PbWO$_{4}$) bars arranged in twelve layers with a total thickness of 27 r.l., sufficiently deep to absorb electron showers totally up to TeV energies.  

\begin{figure}[htbp]
\begin{center}
\includegraphics[width=1.0\textwidth]{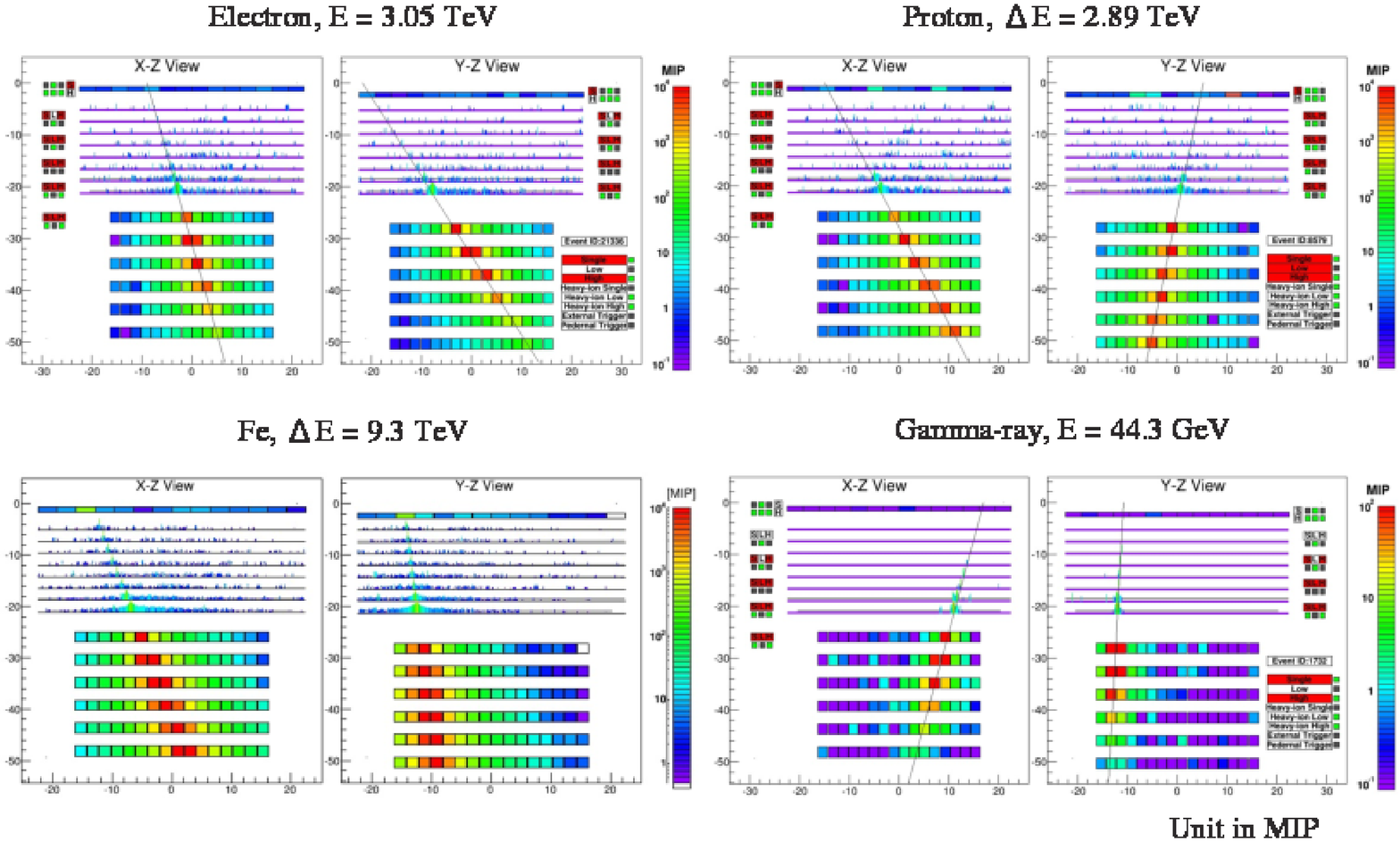}
\caption{\label{fig:CAL_Event_types} Schematic of CAL calorimeter showing CHD, IMC, and TASC subsystems and characteristically identifiable behavior of electrons, protons, nuclei, and gamma rays passing through the detector. Gamma rays are identified by the absence of charge in the CHD and IMC and by the shape and profile of the electromagnetic shower in the TASC. Signal amplitudes are shown according to the right-hand color scale in terms of minimum ionizing particle (MIP) energy deposits. }
\end{center}
\end{figure}

Incident high-energy gamma rays produce Compton electrons and e$^+$e$^-$ pairs in the IMC, resulting in electromagnetic showers.  Gamma ray events are distinguished by the absence of signals in CHD and the top layers of IMC and by the consistency of the observed shower profile in the IMC and TASC with an electromagnetic rather than a hadronic shower. The three CAL subsystems are shown in Fig. \ref{fig:CAL_Event_types} together with examples of the CAL response to electrons, protons, nuclei, and gamma rays.  Incoming gamma ray directions are checked to eliminate events that might have crossed ISS structures in the CAL field of view \citep{Cannady-ICRC2021}. The detector performance is  characterized by Monte Carlo simulations compared in detail to a  series of accelerator calibrations \citep{CALET-energy} and regularly monitored in flight with cosmic ray data.

\begin{figure}[htbp]
\begin{center}
    \hspace{0.5 cm}
    \includegraphics[width=0.6\textwidth]{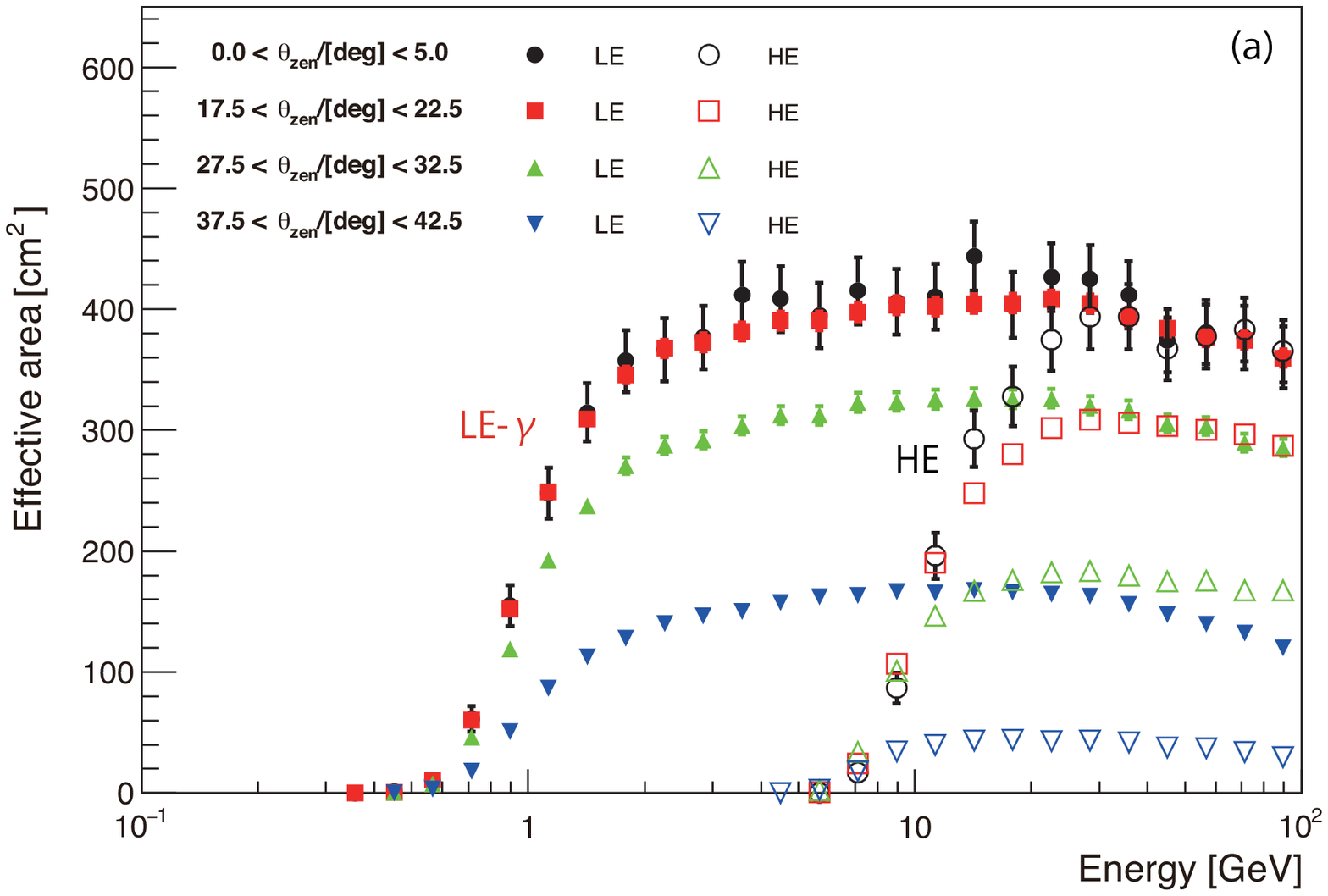}
  \hfill
    \includegraphics[width=0.5\textwidth]{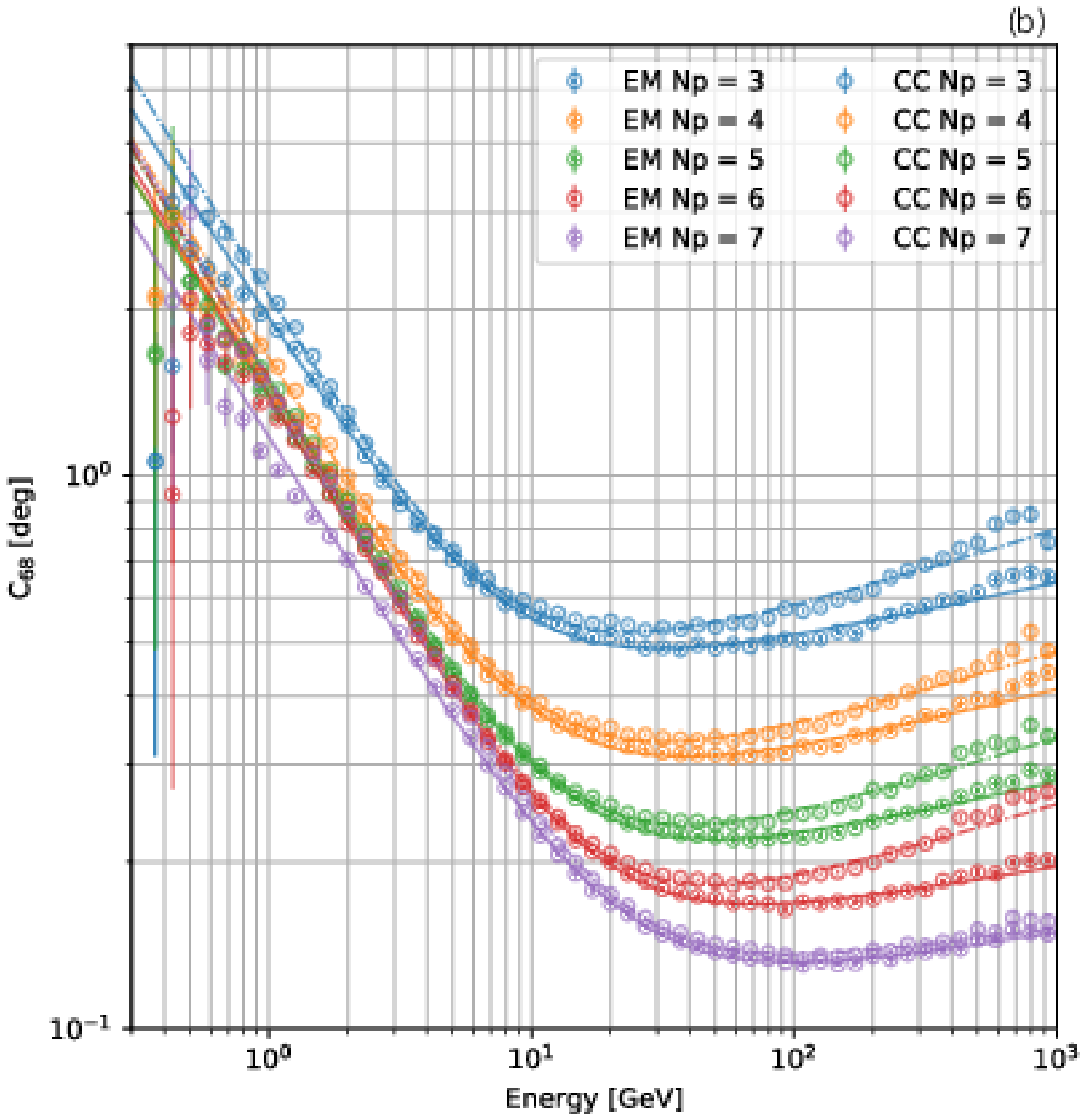}
\caption{\label{CAL_area} a) CAL effective area for gamma rays as a function of energy in four zenith angle ranges for both the LEG (CC Track, solid symbols) and HE (EM Track, open symbols) trigger configurations, from \cite{CALET-O1and2}.  b) CAL angular resolution $C_{68}$ for the two trigger conditions as a function of gamma ray energy and the number of IMC layers used for tracking. Smooth curve shows the fitted resolution function described by \cite{CALET-gammas}.}
\end{center}
\end{figure}

Photon energies are determined by summing the deposited energies in the TASC. At 10 GeV, the energy resolution is $3\%$. The CAL field of view for gamma rays is $\sim 45 ^{\circ}$ from the zenith.

The analysis algorithms (CC Track mode for the LEG trigger and EM Track mode for the HE trigger) are described in \cite{CALET-gammas}.  The effective area is shown for both CC and EM Track mode in  Fig. \ref{CAL_area}a    as a function of gamma ray energy for four separate zenith angle ranges. Effective area reaches $\sim$ 400 cm$^2$ for energies up to $\sim$ 50 GeV, where the identification of zero-charge particles in CHD and IMC begins to be affected by backscatter of higher energy particles.

The angular resolution is shown  in  Fig. \ref{CAL_area}b separately for CC and EM modes as a function of the gamma ray energy and $N_p$, the number of IMC layers used in the tracking. The angular resolution  in Fig. \ref{CAL_area}b is defined as the value $C_{68}$ such that 68$\%$ of simulated events have reconstructed directions deviating from the true direction by an angle $\alpha < C_{68}$.  $C_{68}$ is 0.5$^{\circ}$ or better for all energies above 1 GeV for all but the shortest tracks.   

\begin{figure}[htbp]
\begin{center}
\includegraphics[width=1.0\textwidth]{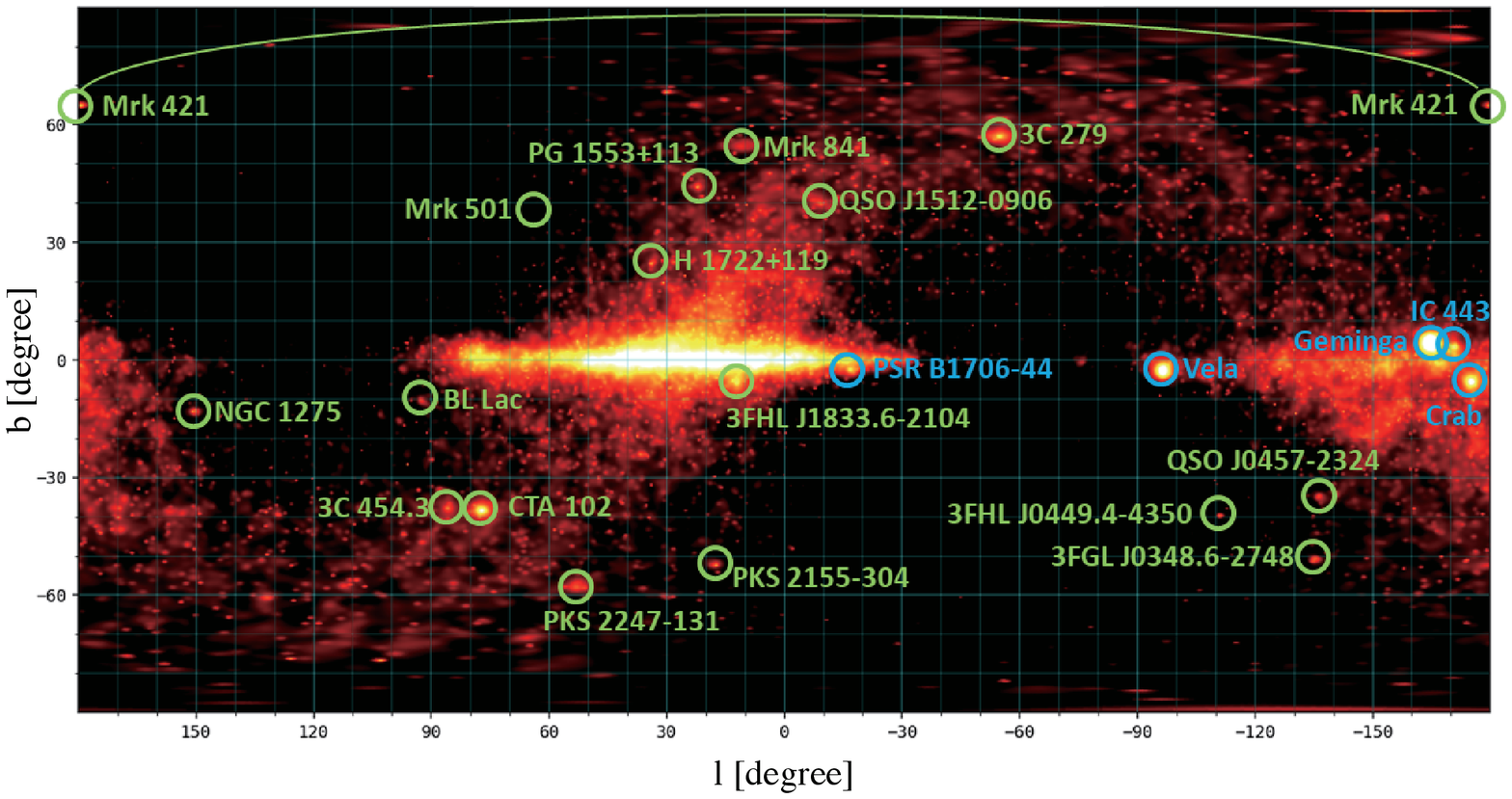}
\caption{\label{fig:CAL_sky_map} Map of sky above 1 GeV observed by CAL in galactic coordinates \citep{Cannady-ICRC2021}. Point-spread
function is determined for each photon event and an order 9 Healpix map ($\sim 7$ arcmin resolution) is filled with the summed smeared signals from each event. Total exposure is accumulated in bins of energy for each
pixel of an order 6 Healpix map ($\sim 55$ arcmin resolution). Resulting counts map is shown in logarithmic scaling.}
\end{center}
\end{figure}

A sky map for 5 years of CAL observations above 1 GeV including  steady gamma ray sources and bright transient events (e.g., CTA102) is shown in Fig. \ref{fig:CAL_sky_map}.  Sources are marked by green or blue circles depending on extragalactic or galactic origin, respectively.  
Although the CAL energy range is far above the typical energy range of GRB prompt emissions, CAL provides the possibility of detecting high energy gamma ray emission from GRBs similar to events  observed up to nearly 100 GeV by {\it Fermi}-LAT \citep{LAT-GRBs}.  

\section{CALET search for GW events during the LIGO/Virgo Third observing run}
CALET was still in its commissioning phase at the time of the initial LIGO detection of GW 150904. Since October 2015, however, CALET has actively searched for  electromagnetic counterparts of gravitational wave events. The CGBM and CAL searches for Advanced LIGO and Advanced Virgo events during O1 and O2 have been described by \cite{Yamaoka-2017} and   \cite{CALET-O1and2} respectively. Here we describe the CALET search since the start of O3 in 2019.  The LIGO/Virgo collaboration (LVC) reported 56 gravitational events (not including  retracted events) in O3.  Also, LVC and {\it Fermi}-GBM reported one sub-threshold event \citep{GBM190816}. The 57 events are shown in Table \ref{tab:o3_events}.  All information in Table \ref{tab:o3_events} is based on GCN circulars and The Gravitational-Wave Candidate Event Database (GraceDB, https://gracedb.ligo.org/) operated by LIGO.   Here `Possible source' shows the most probable source as listed in GraceDB. CALET has searched for electromagnetic counterparts of gravitational waves in both the CGBM and CAL data and results   have been reported in GCN circulars  (Table \ref{tab:o3_events}).  We describe the details of the CGBM  and CAL analyses separately in the following two subsections. 

\subsection{CGBM analysis}
As of the end of November 2021, CGBM has detected  271 GRBs, with 12 \% of the CGBM GRBs  classified as short GRBs \citep{CALET-GRBs-2021}.
CGBM observations of the O3 events are summarized in Table \ref{tab:cgbm_obs}.  Here $T_{\mathrm 0}$ is the time of the gravitational wave event reported in GraceDB and listed in Table 2.
Since the duty cycle of CGBM is $\sim$~60 \% due mainly to passages through the SAA, CGBM high voltages were sometimes turned off at gravitational wave trigger times.  The high voltage status at the trigger time of each gravitational wave event was recorded in the housekeeping data and the status of the high voltages is shown in Table \ref{tab:cgbm_obs} under ‘CGBM trigger’: Here `Disabled' means the on-board trigger system was disabled at $T_{\mathrm 0}$ either because the CGBM high voltages were off or the CGBM event buffer was full. `No trigger' means the on-board trigger system was not triggered during the time interval $T_{\mathrm 0} \pm 60$~s even if the on-board trigger system was enabled.  
The monitor data were inspected on the ground for each O3 event to confirm that no potential candidate event occurred within $T_{\mathrm 0} \pm 60$~s.
For each event, the summed LIGO/Virgo localization probability above the CGBM horizon was calculated using the LIGO/Virgo sky maps from GraceDB.  If the summed  LIGO/Virgo localization probability above the horizon was greater than 1\%, a targeted signal search was then performed using the TH data.  Light curves were constructed from the TH data for each CGBM detector with 0.125 s time binning in twenty energy bands for $T_{\mathrm 0} - 60$ s to $T_{\mathrm 0} + 60$ s.  SNR was recalculated  for each light curve with an expanded set of search parameters (Table \ref{tab:lc_cond})  to search for a significant signal within $T_{\mathrm 0} \pm 60$s. In the ground analysis, $\Delta t_{\mathrm{BG}}$ was taken from both sides of $\Delta t$ with one exception:  In the case of the high voltage turning on or off within $\pm$ 60 s of the trigger time, $\Delta t_{\mathrm{BG}}$ was taken only from the period when the high voltage was on.  SNR was therefore calculated every 0.125 s for 1440 separate conditions summarized in Table \ref{tab:lc_cond}.

$\sigma_{{\rm max}, T_{0}}$ in Table \ref{tab:cgbm_obs} shows the highest SNR from the set of SNRs calculated for all conditions such that the foreground intervals $\Delta t$ include $T_{\mathrm 0}$.  `Conditions for $\sigma_{{\rm max}, T_{0}}$' shows the conditions resulting in $\sigma_{{\rm max}, T_{0}}$.  $\sigma_{{\rm max}, {\rm 60s}}$ shows the highest SNR calculated for all conditions in $T_{\mathrm 0} \pm 60$~s.  `Conditions for $\sigma_{{\rm max}, {\rm 60s}}$' shows the conditions corresponding to $\sigma_{{\rm max}, {\rm 60s}}$.  Finally, $T_{\rm max}$ is  the start time of the foreground interval relative to $T_{\mathrm 0}$ when the SNR was equal to $\sigma_{{\rm max}, {\rm 60s}}$.

\begin{figure}[htbp]
\hspace*{2.5cm}
\begin{center}
\includegraphics[width=0.7\textwidth]{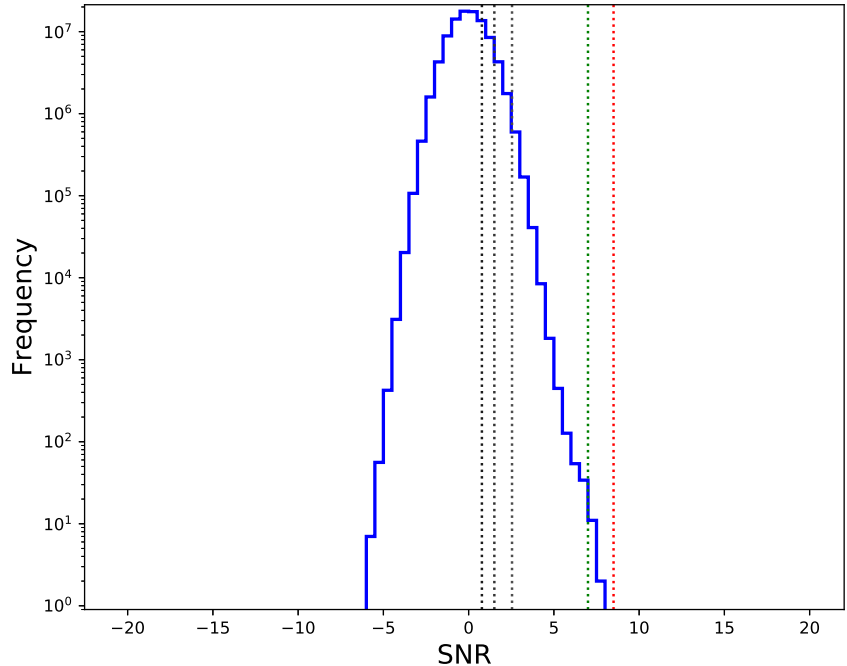}
\caption{Histogram of SNRs calculated for SGM high gain data for 2019/10/01. Vertical dashed lines correspond to (from left to right) 68th percentile, 90th percentile, 99th percentile,  $7 \sigma$, and $8.5 \sigma$.} \label{fig:snr_hist}
\end{center}
\end{figure}

Background varies over an orbit mainly due to the variation of the trapped charged particle flux and activity of bright X-ray sources in the field of view.
Since the estimation of background counts in the SNR calculation is based on the summation of the observed counts before and after the foreground time interval, estimated background counts during $\Delta t$ are sometimes underestimated or overestimated depending on the background variation during the orbit, affecting the calculated SNRs.  The distribution of SNRs is not described by a normal distribution: As an example, Figure \ref{fig:snr_hist} shows the distribution of SNRs calculated using SGM High Gain data for one day.  The histogram includes SNRs calculated every 0.125 s continuously with all conditions and gain settings. Individual bins are therefore not independent of each other.  As seen in the figure, the distribution extends up to SNR $\sim7$.  On this particular day, a small tail is present at high SNR due to an increased counting rate in two successive orbits at  high latitude. Including this contribution of high-SNR events due to low-energy charged particles, the fraction of events with SNR $>7$ is 1.6 $\times 10^{-7}$. 
We set the threshold for the off-line trigger at SNR = 7 and require that candidate events show up in multiple detectors (HXM1, HXM2, and SGM) and multiple energy channels, that the event arrival time falls within $T_{\mathrm 0} \pm 60$s, and that CAL CHD and IMC see no simultaneous increase in the low-energy charged particle rate.  

As a check on the reasonableness of these event selection criteria, we have performed a search for events simulating gravitational wave counterparts by searching for events either one orbit prior to or one orbit after the times of the 57 events listed in Table 2 -- i.e., at times and positions where the instrument is at approximately the same latitude and in approximately the same pointing direction as at the time of the actual gravitational wave event. One event was seen with SNR = 7.13, but only in HXM2 -- i.e., no multi-channel events were found that successfully masqueraded as candidate CGBM counterpart events. 

\begin{figure}[htbp]
\begin{center}
  \begin{minipage}[b]{0.49\textwidth}
    \includegraphics[width=\textwidth]{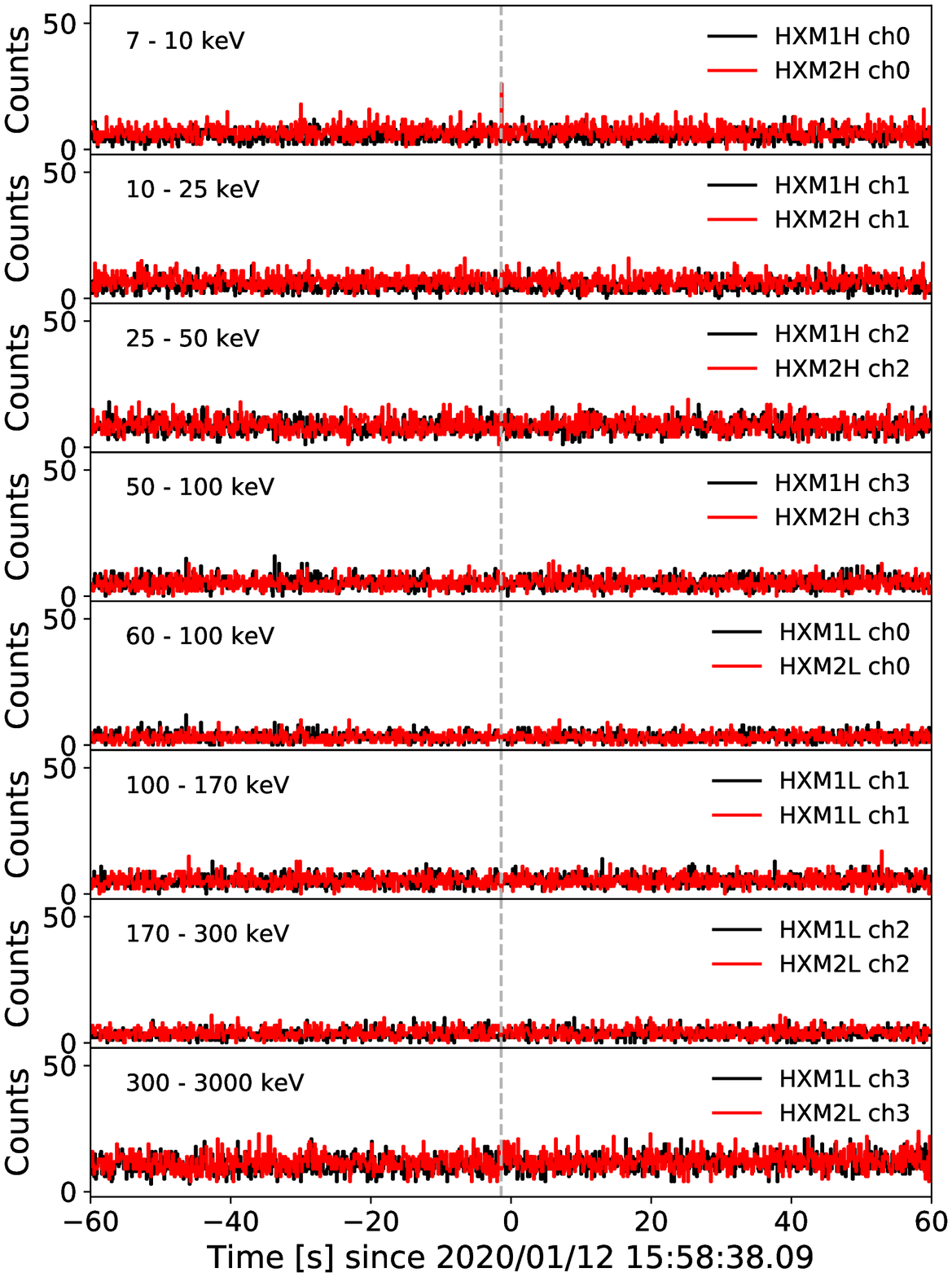}
  \end{minipage}
  \hfill
  \begin{minipage}[b]{0.49\textwidth}
    \includegraphics[width=\textwidth]{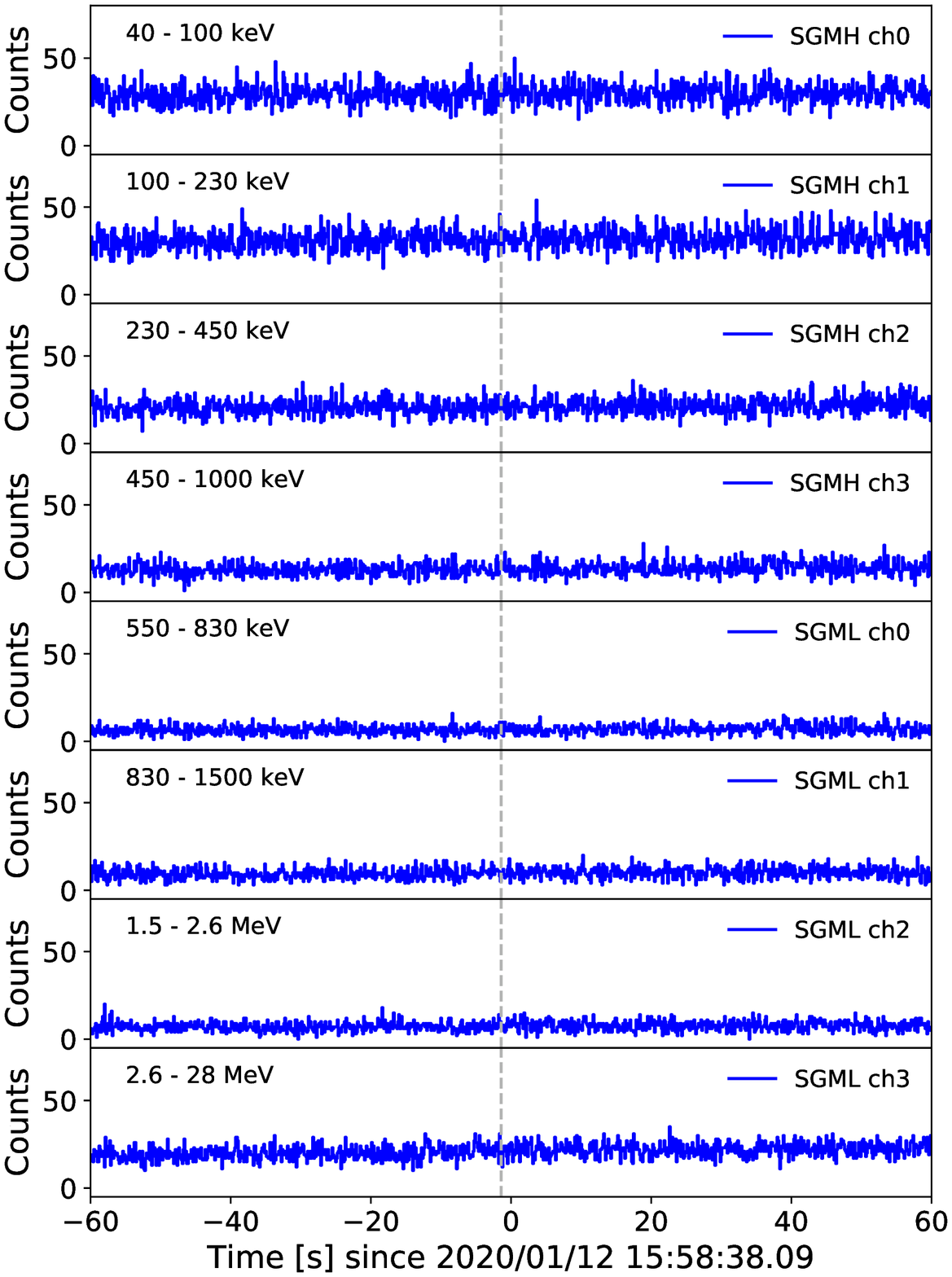}
  \end{minipage}
\caption{\label{fig:S200112r} Time histories of counts detected by CGBM within $\pm$ 60 s of LIGO/Virgo event S200112r. Dashed lines show time $T_{\mathrm 0} - 1.41$ prior to LIGO/Virgo event S200112r. CGBM signal excess is seen only in the lowest energy channel of HXM2.}
\end{center}
\end{figure}

In the time windows $T_{\mathrm 0} \pm 60$~s corresponding to real LIGO/Virgo events, one LIGO/Virgo event (S200112r) was found with a nearby CGBM SNR $>7$ (Fig. \ref{fig:S200112r}).
In this case, the time of highest SNR ($T_{\rm max}$) was $T_{\mathrm 0} - 1.41$~s. However, the signal can be seen in only the lowest energy  channel of HXM2; HXM1 and SGM detected no significant signal. We conclude that this signal is likely a random fluctuation not likely to be physically associated with S200112r.

Since no acceptable GRB candidate associated with any of the LIGO/Virgo GW events was found in the CGBM data, we estimated upper limits for the X-ray/gamma ray flux.  For each GW event, we simulated typical GRB spectra in the TH data using the `fakeit' command of XSPEC version 12.10.1 (https://heasarc.gsfc.nasa.gov/docs/xanadu/xspec/issues/archive/issues.12.10.1s.html and \cite{XSPEC}) assuming typical spectra and durations for short GRBs.  We used the CGBM response matrix database which includes CGBM response files calculated for 5$^{\circ}$ pitch in zenith angle and azimuth.  
For the assumed input spectrum, we used a Band function and a power law with exponential cutoff and standard values for photon indices and $E_{\mathrm{peak}}$ characteristic of short GRBs: For the Band function, we used $\alpha = -0.46$, $\beta = -2.98$, and $E_{\mathrm{peak}}=413$ keV \citep{Poolakkil} and for the cutoff power law we used $\alpha = -0.42$ and $E_{\mathrm{peak}}=566$ keV  \citep{GBM-Racusin,  GBM-170817}. We assume a burst duration of 1 second.
Based on the LIGO/Virgo sky maps, we took the direction of the source to be that 
direction within the CGBM field of view for which the localization probability was maximum and applied the relevant CGBM response matrices. Tables 5 - 7 show the resulting time-averaged flux upper limits at the level of 7 $\sigma$ calculated separately for HXM1, HXM2, and SGM in the 10~keV - 1 MeV energy range. Here $P$ corresponds to the summed probability in the field of view of each detector (HXM1, HXM2, or SGM), $\alpha$ and $\delta$ are the highest probability directions in Equatorial coordinates of the GW sources in the field of view of each detector, and $\theta$ and $\phi$ are the zenith and azimuth angles of incident photons striking the detector from the direction of the GW source. (Since the CGBM angular response is calculated on a $5^{\circ}$ grid, $\theta$ and $\phi$ are tabulated with $5^{\circ}$ precision.)

\subsection{CAL analysis} 
Results of the CAL observations of the 57 events reported by LIGO/Virgo for O3 are shown in Table \ref{tab:cal_obs}. Of the 57 events, 20 were in the CAL field of view. Of those, 13 occurred while CAL was in HE mode and 7 occurred when CAL was in LEG mode. In no case did CAL detect any events from the allowed region within $\pm$ 60 s of the LIGO/Virgo $T_{\mathrm 0}$.  The pointing direction of the center of the CAL field of view is given in the columns labelled $\alpha$ and $\delta$. "Coverage" is the fraction of the overlapping region of the LIGO/Virgo localization map covered by the CAL field of view during the interval  $T_{\mathrm 0} \pm 60$ s. 

Based on the LIGO-Virgo sky maps  and taking ISS structures in the field of view into account for each event, we calculate an effective area for each direction as a function of energy. 
CAL observations of the GeV sky reflect a combination of individual galactic and extragalactic sources together with both galactic and extragalactic diffuse emission in good agreement with {\it Fermi}-LAT Pass 8 observations \citep{CALET-gammas,Cannady-ICRC2021}. Based on this, an expected number of background events is calculated for each time interval and direction in the CAL search for counterparts; typically, the expected $N_{\mathrm{BG}} \sim 0.1$ or less. In the case of a null event, we assume a power law spectrum for a potential gamma ray burst with a single power law photon index of -2 \citep{Poolakkil}, taking into account the CAL sensitivity as a function of energy,  and estimate an upper limit (90~\% confidence level) on the gamma ray flux based on 2.44 events above expected background. Assuming a burst duration $\Delta t = 1$s, upper limits are calculated for the time interval $T_{\mathrm 0} \pm 60$ s in units of erg cm$^{-2}$ s$^{-1}$ for the energy range 10 - 100 GeV (with the CAL in HE mode) or 1 - 10 GeV (with the CAL  in LEG mode).   The maximum time-averaged flux for an individual pixel in the LIGO/Virgo localization area is listed as the upper limit given in the final column of Table 8.

\begin{figure}[htbp]
\begin{center}
\includegraphics[width=0.8\textwidth]{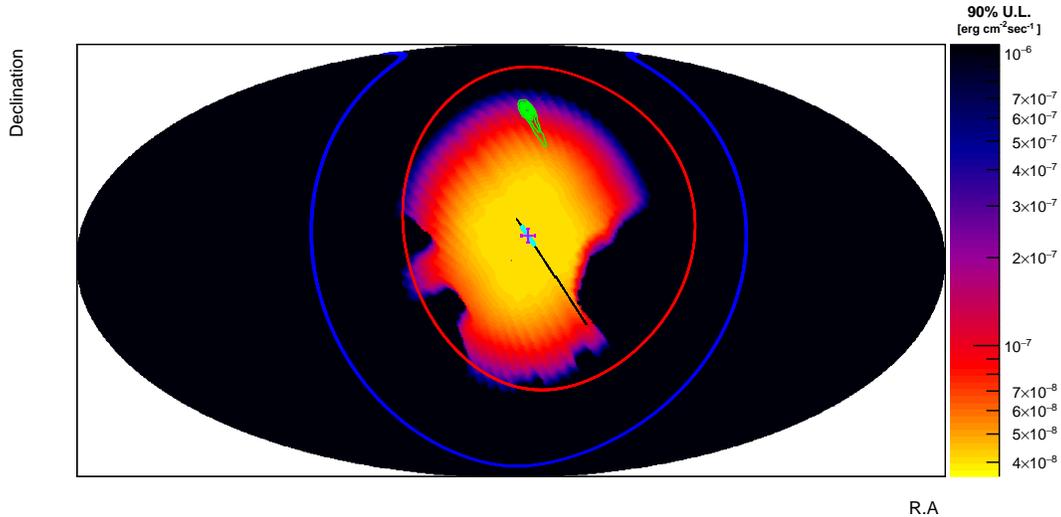}
\caption{\label{190408an} 90 \% confidence level upper limits observed by CAL in the energy range 1 - 10 GeV during the interval $\pm 60$ s around the time of GW 190408an reported by LIGO/Virgo. Intensity scale is given in units of ergs cm$^{-2}$ s$^{-1}$. Red and blue circles are the HXM and SGM fields of view, respectively.}
\end{center}
\end{figure}

The lowest upper limit in Table 8 is for the CAL observation of the sky at the time of S190408an. No CGBM on-board trigger occurred near the GW event time; no excess count rate was seen in the ground analysis of the HXM and SGM data within 60 s of $T_{\mathrm 0}$; and no CAL gamma ray events were detected within 60 s of the GW trigger time.  Fig. \ref{190408an} shows the map of 90\% confidence level upper limits measured by CAL during the time interval $T_{\mathrm 0} - 60$ s to $T_{\mathrm 0} + 60$ s. The pointing direction of the CAL during the observation is marked by the cyan contour extending upward from the lower right to the middle of the diagram. The magenta cross marks the zenith direction at $T_{\mathrm 0}$. The green area near the extreme upper edge of the CAL 90\% upper limit region marks the localization contours reported by LIGO/Virgo.  The red and blue circles are the HXM and SGM fields of view ignoring effects of the ISS structures at $T_{\mathrm 0}$, respectively.

\section{Discussion and conclusions}

The neutron star merger event GRB 170817A was detected by {\it Fermi}-GBM and {\it INTEGRAL} $\sim 1.7$ s after the gravitational wave event at a 10 - 1000 keV flux  level (based on the {\it Fermi}-GBM measurement) of $(5.5 \pm 1.2) \times 10^{-7}$ ergs cm$^{-2}$ s$^{-1}$  \citep{GBM-170817}. The observed properties of the gamma ray event were those of an ordinary although 
sub-luminous short GRB. Previously, {\it Fermi}-GBM also reported a $2.9 \sigma$ signal from a candidate short GRB counterpart GW 150914-GBM seen 0.4 s after the gravitational wave event. The reported   fluence level  over 1 s was $ (2.4 - 2.8) \times 10^{-7}$ ergs cm$^{-2}$ depending on whether a power law or Comptonized model was used to fit the data  \citep{Connaughton}. Given the low significance and lack of confirmation by other instruments, GBM did not claim this event as a real counterpart to GW 150914. 
Since models of neutron star-neutron star and neutron star-black hole mergers do not provide strong constraints on the expected X-ray/gamma ray fluxes \citep{Rees1994,Rees2005,Phinney2009,Rosswog2015,Fernandez2016}, these two {\it Fermi}-GBM/{\it INTEGRAL} events suggest that a 10 - 1000 keV sensitivity level on the order of several times $10^{-7}$ erg cm$^{-2}$ s$^{-1}$ is a desirable target flux for a hard X-ray/$\gamma$-ray counterpart search.
The $7 \sigma$ flux limits (averaged over 1 s) listed for CGBM in Tables 5-7 are typically factors of a few times higher than this, largely due to the larger collecting power of GBM compared to CGBM.  Nevertheless, there are several possible reasons why GBM or other detectors might miss a real event: The intersection of the LIGO/Virgo localization probability map with the $\gamma$-ray detector FOV may be too low; the event may be beamed in an unfavorable direction; or the $\gamma$-ray detector may be disabled or experiencing high background as it passes through a high-latitude region or the SAA. Having multiple detectors monitoring for counterparts is therefore essential in order to search effectively for rare events like GW 170817-GRB 170817A.

{\it Fermi}-LAT searches for GeV gamma rays in coincidence with LIGO/Virgo events \citep{LAT-150914,GBM-Racusin,LAT-170817} have typically been at 95$\%$ confidence sensitivity levels of $(3 - 5) \times 10^{-10}$ ergs cm$^{-2}$ s$^{-1}$ over the energy range 0.1 - 1 GeV. Soft GRBs observed by LAT are often delayed and have longer durations than the prompt signals \citep{LAT-GRBs}, consistent with an afterglow origin \citep{afterglow-DePasquale,afterglow-Gehrels, afterglow-Kouveliotou}. LAT counterpart searches have therefore looked for excesses on time scales  up to 10 days before and after $T_{\mathrm 0}
$. In two cases, CAL has detected GeV gamma ray candidates delayed by 105 and 244 s and from within $0.6^{\mathrm o}$ and $1.3^{\mathrm o}$ of a GRB recorded by CGBM \citep{CALET-GRBs-2021}. However, the CAL energy range is 1-10 GeV (for LEG mode) or 10-100 GeV (for HE mode), where fluxes are expected to be lower than in the {\it Fermi}-LAT range. In addition, the smaller CAL telescope will not provide as much sensitivity as LAT for delayed emission, and so the present CAL counterpart search is limited to the prompt emission, where a favorable pointing direction may provide sensitivity greater than that of other larger instruments. As discussed above, CAL has detected no candidate events. The resulting upper limits in Table 8 range from $3.0 \times 10^{-7}$ to $4.8 \times 10^{-4}$ ergs cm$^{-2}$ s$^{-1}$.
\\

CALET is currently approved  to continue operating through 2024. We gratefully acknowledge JAXA’s contributions to the development of CALET and to the operations aboard
the JEM-EF on the International Space Station. We also wish to express our sincere gratitude to Agenzia Spaziale Italiana (ASI) and NASA for their support of
the CALET project. 
In Japan, this work was supported in part by JSPS KAKENHI Grant Numbers 26220708, 19H05608, 17H02901, 21K03592, and 20K22352 and
by the MEXT-Supported Program for the Strategic Research Foundation at Private Universities (2011-2015) (No.S1101021) at Waseda University. 
The CALET effort in Italy is supported by ASI under agreement 2013-018-R.0 and its amendments. 
The CALET effort in the United States is supported by NASA through Grants 80NSSC20K0397, 80NSSC20K0399, and NNH18ZDA001N-APRA18-000.  
A part of this research is made possible by use of data obtained from DARTS at ISAS/JAXA.


\newpage\begin{longtable}{lcccc} 
\caption{Summary of LVC gravitational wave events in O3 and CALET follow-up observations.}
\label{tab:o3_events}\\
\hline\hline
Event name  & Possible source & Event time ($T_{0}$) & \ LVC GCN circular \# \ &\ CALET GCN circular \#\\ \hline

%\centering
%\tiny
%\small
%\hspace{-2.0cm}
\endhead
    S190408an   & BBH ($>$99 \%) &2019/04/08 18:18:02.288180  & 24069 & 24088\\
    S190412m    & BBH ($>$99 \%) & 2019/04/12 05:30:44.165622  & 24098 & - \\
    S190421ar   & BBH (97 \%) & 2019/04/21 21:38:56.250977   & 24141, 24375 & - \\
    S190425z    & BNS ($>$99 \%) & 2019/04/25 08:18:05.017147   & 24168, 24228 & 24218\\
    S190426c    & Terrestrial (58 \%) &2019/04/26 15:21:55.336540   & 24237, 24277, 24279, 24411 & 24276\\
    S190503bf   & BBH (96 \%) & 2019/05/03 18:54:04.294490   & 24377 & 24403\\
    S190510g    & Terrestrial (58 \%) & 2019/05/10 02:59:39.291636   &24442, 24448, 24462, 24489 & 24495\\
    S190512at   & BBH (99 \%) & 2019/05/12 18:07:14.422363   & 24503, 24584 & 24531\\
    S190513bm   & BBH (94 \%) & 2019/05/13 20:54:28.747089 & 24522 & 24548\\
    S190517h    & BBH (98 \%) & 2019/05/17 05:51:01.830582   & 24570 & 24593\\
    S190519bj   & BBH (96 \%) & 2019/05/19 15:35:44.397949   & 24598 & 24617\\
    S190521g    & BBH (97 \%) & 2019/05/21 03:02:29.447266   & 24621, 24640 & 24648\\
    S190521r    & BBH ($>$99 \%) & 2019/05/21 07:43:59.463379    & 24632 & 24649\\
    S190602aq   & BBH ($>$99 \%) & 2019/06/02 17:59:27.089355  & 24717 & 24735\\
    S190630ag   & BBH (94 \%) & 2019/06/30 18:52:05.179550  & 24922, 25094 & 24960\\
    S190701ah   & BBH (93 \%) & 2019/07/01 20:33:06.577637  & 24950, 24987 & 24970\\
    S190706ai   & BBH (99 \%) & 2019/07/06 22:26:41.344727  & 24998, 25049 & 25027\\
    S190707q    & BBH ($>$99 \%) & 2019/07/07 09:33:26.181226  & 25012, 25048 & 25033\\
    S190718y    & Terrestrial (98 \%) & 2019/07/18 14:35:12.067865  & 25087, 25107 & 25099\\
    S190720a    & BBH (99 \%) & 2019/07/20 00:08:36.704102  & 25115, 25138 & 25134\\
    S190727h    & BBH (92 \%)  & 2019/07/27 06:03:33.985887  & 25164, 25249 & 25184\\
    S190728q    & MassGap (52 \%) & 2019/07/28 06:45:10.529205  & 25187, 25208 & 25214\\
    S190814bv   & NSBH ($>$99 \%) & 2019/08/14 21:10:39.012957  & 25324, 25333 & 25390\\
    Fermi GBM-190816 & sub-threshold& 2019/08/16 21:22:13.027 & 25406 & -\\
    S190828j    & BBH ($>$99 \%) & 2019/08/28 06:34:05.756472  & 25497, 25504, 25861 & 25536\\
    S190828l    & BBH ($>$99 \%) & 2019/08/28 06:55:09.886557  & 25503, 25782 & 25537\\
    S190901ap   & BNS (86 \%) & 2019/09/01 23:31:01.837767  & 25606, 25614 & 25647\\
    S190910d    & NSBH (98 \%) & 2019/09/10 01:26:19.242676  & 25695, 25723 & 25734\\
    S190910h    & BNS (61 \%) & 2019/09/10 08:29:58.544448  & 25707, 25778 & 25735\\
    S190915ak   & BBH ($>$99 \%) & 2019/09/15 23:57:02.690891 & 25753, 25773 & 25770\\
    S190923y    & NSBH (68 \%) & 2019/09/23 12:55:59.645508  & 25814 & 25830\\
    S190924h    & MassGap ($>$ 99 \%) & 2019/09/24 02:18:46.846654  & 25829, 25905, 25909 & 25844\\
    S190930s    & MassGap (95 \%) & 2019/09/30 13:35:41.246810  & 25871, 25968 & 25891\\
    S190930t    & NSBH (74 \%) & 2019/09/30 14:34:07.685342  & 25876 & 25892\\
    S191105e    & BBH (95 \%) & 2019/11/05 14:35:21.933105  & 26182, 26245 & 26195\\
    S191109d    & BBH ($>$99 \%) & 2019/11/09 01:07:17.220703  & 26202 & 26236\\
    S191129u    & BBH ($>$99 \%) & 2019/11/29 13:40:29.197372  & 26303, 26383 & 26321\\
    S191204r    & BBH ($>$99 \%) & 2019/12/04 17:15:26.091822  & 26334 & 26358\\
    S191205ah   & NSBH (93 \%) & 2019/12/05 21:52:08.568738  & 26350 & 26377\\
    S191213g    & BNS (77 \%) & 2019/12/13 04:34:08.142224  & 26402, 26417 & 26419\\
    S191215w    & BBH ($>$99 \%) & 2019/12/15 22:30:52.333152 & 26441, 26518 & 26465\\
    S191216ap   & BBH ($>$99 \%) & 2019/12/16 21:33:38.472999 & 26454, 26505, 26570 & 26481\\
    S191222n    & BBH ($>$99 \%) & 2019/12/22 03:35:37.119478  & 26543, 26572 & 26602\\
    S200105ae   & Terrestrial (97 \%) & 2020/01/05 16:24:26.057208 &26640, 26688 & 26664\\
    S200112r    & BBH ($>$99 \%) & 2020/01/12 15:58:38.093931 & 26715 & 26740\\
    S200114f    & - & 2020/01/14 02:08:18.239300 & 26734 & 26761\\
    S200115j    & MassGap ($>$99 \%) & 2020/01/15 04:23:09.742047 &26759, 26807 & 26797\\
    S200128d    & BBH (97 \%) & 2020/01/28 02:20:11.903320 & 26906 & 26924\\
    S200129m    & BBH ($>$99 \%) & 2020/01/29 06:54:58.435104 & 26926 & 26941\\
    S200208q    & BBH ($>$99 \%) & 2020/02/08 13:01:17.991118 & 27014, 27036 & 27030\\
    S200213t    & BNS (63 \%) & 2020/02/13 04:10:40.327981 & 27042, 27092, 27096 & 27084\\
    S200219ac   & BBH (96 \%) & 2020/02/19 09:44:15.195312 & 27130, 27214 & 27149\\
    S200224ca   & BBH ($>$99 \%) & 2020/02/24 22:22:34.405762 & 27184, 27262 & 27231\\
    S200225q    & BBH (96 \%) & 2020/02/25 06:04:21.396973 & 27193, 27229 & 27232\\
    S200302c    & BBH (89 \%) & 2020/03/02 01:58:11.519119 & 27278, 27292 & 27299\\
    S200311bg   & BBH ($>$99 \%) & 2020/03/11 11:58:53.397788 & 27358, 27382 & 27372\\
    S200316bj   & MassGap ($>$99 \%) & 2020/03/16 21:57:56.157221 & 27388, 27419 & 27405\\
    
\hline 
\end{longtable}

\newpage
\begin{longtable}{lccccccc}
\caption{Summary of CGBM observations for gravitational wave events in O3.}
\label{tab:cgbm_obs}\\
\hline\hline
    Event name  &  CGBM trigger & $P_{\rm h}$ & $\sigma_{{\rm max}, T_{0}} $ &  Conditions  & $\sigma_{{\rm max}, {\rm 60s}}$ & Conditions  & $T_{\rm max}$ \\
    & & & & for $\sigma_{{\rm max}, T_{0}} $ & & for $\sigma_{{\rm max}, {\rm 60s}}$ &   \\
    \hline
\endhead
    
    S190408an   &No trigger& 100 \%  & 4.70 &HXM2, Low, ch0-2& 5.41 & HXM1, Low, ch2 & -37.61\\
    &&&&$\Delta t$=4s, $\Delta t_{\mathrm{BG}}$=16s&&$\Delta t$=0.25s, $\Delta t_{\mathrm{BG}}$= 8s&\\
    S190412m    & Disabled & - &-&-&-&-&- \\
    
    S190421ar   & No trigger& 0 \% &-&-&-&-&- \\
    
    S190425z    & Disabled & - &-&-&-&-&- \\
    
    S190426c    & Disabled & - &-&-&-&-&- \\
    
    S190503bf   & Disabled & - &-&-&-&-&- \\
    
    S190510g    & No trigger & 16 \% & 3.95 &SGM, High, ch1-3 & 5.22 & HXM1, High, ch1 & -17.00\\
    &&&&$\Delta t$=4s, $\Delta t_{\mathrm{BG}}$=8s&&$\Delta t$=0.125s, $\Delta t_{\mathrm{BG}}$= 64s&\\
    
    S190512at   & No trigger & 100 \% & 4.11 & SGM, High, ch0-3 & 5.22 &  HXM2, High, ch3 & 28.96 \\
    &&&&$\Delta t$=4s, $\Delta t_{\mathrm{BG}}$= 8s&&$\Delta t$=0.125s, $\Delta t_{\mathrm{BG}}$= 16s&\\

    S190513bm   & No trigger & 100 \% & 4.41 &  HXM2, High, ch2-3 & 5.34 & SGM, Low, ch0-2 & 10.26\\
    &&&& $\Delta t$=4s, $\Delta t_{\mathrm{BG}}$=16s&&$\Delta t$=4s, $\Delta t_{\mathrm{BG}}$=8s&\\
    
    S190517h    & No trigger & 86 \% & 3.99 & HXM1, Low, ch0& 5.47 & SGM, Low, ch2-3 & 4.04 \\
    &&&&$\Delta t$=1s, $\Delta t_{\mathrm{BG}}$=8s &&$\Delta t$=4s, $\Delta t_{\mathrm{BG}}$=64s&\\

    S190519bj   & No trigger & 100 \% & 4.54 & HXM2, High, ch2 & 5.69 & HXM1, Low, ch0 &-7.15 \\
    &&&& $\Delta t$=4s, $\Delta t_{\mathrm{BG}}$=8s&& $\Delta t$=0.125s, $\Delta t_{\mathrm{BG}}$= 8s&\\
    
    S190521g    & Disabled &-&-&-&-&- &- \\
    S190521r    & Disabled &-&-&-&-&- &-\\
    
    S190602aq   & No trigger & 99 \% & 4.56 & HXM2, Low, ch1 & 5.22 & HXM2, High, ch3& -41.01\\
    &&&&$\Delta t$=4s, $\Delta t_{\mathrm{BG}}$=8s&&$\Delta t$=1s, $\Delta t_{\mathrm{BG}}$=32s &\\

    S190630ag   & Disabled &-&-&-&-&- & -\\
        
    S190701ah   & No trigger & 19 \%& 4.80 & HXM2, Low, ch2& 5.43 & HXM1, High, ch0& -19.20\\
    &&&&$\Delta t$=4s, $\Delta t_{\mathrm{BG}}$= 8s&&$\Delta t$=0.125s, $\Delta t_{\mathrm{BG}}$=16s&\\
    
    S190706ai   & Disabled & - & - &-&-&-&-  \\
    
    S190707q    & No trigger & 76 \% & 3.87 & SGM, High, ch1-3 & 5.13 & HXM2, Low , ch2& 10.60\\
    &&&& $\Delta t$=4s, $\Delta t_{\mathrm{BG}}$=32s&& $\Delta t$=0.125s, $\Delta t_{\mathrm{BG}}$=8s&\\
    
    S190718y    & No trigger & 22 \% & 3.54 & HXM2, Low, ch0& 5.13 & HXM2, Low, ch3 & 20.22\\
    &&&&$\Delta t$=4s, $\Delta t_{\mathrm{BG}}$=8s&&$\Delta t$=4s, $\Delta t_{\mathrm{BG}}$=8s&\\
    
    S190720a    & Disabled &-&-&-&-&- & -\\
        
    S190727h    & No trigger & 14 \%& 4.04 & HXM1, High, ch2 & 5.66 & HXM2, Low, ch1& -34.82\\
    &&&& $\Delta t$=0.25s, $\Delta t_{\mathrm{BG}}$=16s&& $\Delta t$=0.25s, $\Delta t_{\mathrm{BG}}$=8s&\\
    
    S190728q    & No trigger & 0 \% &-&-&-&-&- \\
    S190814bv   & Disabled & - &-&-&-&-&-  \\
    
    Fermi GBM-190816 & No trigger & 66 \% & 3.78 & HXM2, High ch1 & 5.25 & HXM1, Low, ch2 & -18.97\\
    &&&& $\Delta t$=2s, $\Delta t_{\mathrm{BG}}$=8s &&  $\Delta t$=1s, $\Delta t_{\mathrm{BG}}$=8s &\\
     
    S190828j    & No trigger & 28 \% & 3.33 & HXM2, High, ch0-3 &5.31 & HXM1, High, ch0-1& 26.45\\
    &&&& $\Delta t$=0.25s, $\Delta t_{\mathrm{BG}}$=64s&& $\Delta t$=4s, $\Delta t_{\mathrm{BG}}$=8s&\\
        
    S190828l    & No trigger & 79 \% & 3.36 & SGM, Low, ch0& 4.82 & HXM2, High, ch1-2 & -47.17\\
    &&&& $\Delta t$=2s, $\Delta t_{\mathrm{BG}}$=8s&& $\Delta t$=0.25s, $\Delta t_{\mathrm{BG}}$=8s&\\
    
    S190901ap   & Disabled & 82 \% & 3.94 & SGM, High, ch0-2 &  5.72 & SGM, Low, ch1 & 24.02\\
    &&&& $\Delta t$=4s, $\Delta t_{\mathrm{BG}}$=8s&& $\Delta t$=4s, $\Delta t_{\mathrm{BG}}$=8s&\\
    
    S190910d    & No trigger & 77 \% & 5.59 & HXM1, High, ch0-1 & 6.31 & HXM1, Low, ch1-3 & -42.07\\
    &&&& $\Delta t$=4s, $\Delta t_{\mathrm{BG}}$=8s&& $\Delta t$=4s, $\Delta t_{\mathrm{BG}}$=8s&\\
    
    S190910h    & No trigger & 78 \% & 3.84 & SGM, Low, ch1 & 6.57 & HXM1, Low, ch2 & 3.71\\
    &&&& $\Delta t$=4s, $\Delta t_{\mathrm{BG}}$=8s&& $\Delta t$=0.25s, $\Delta t_{\mathrm{BG}}$=8s&\\
        
    S190915ak   & No trigger & 100 \% & 4.62 & HXM1, High, ch0 & 5.47 & HXM2, High, ch1-2 & 32.54\\
    &&&& $\Delta t$=2s, $\Delta t_{\mathrm{BG}}$=8s&& $\Delta t$=0.125s, $\Delta t_{\mathrm{BG}}$=64s&\\
    
    S190923y    & No trigger & 68 \% & 4.19 & HXM1, High, ch2 & 5.06 & HXM1, Low, ch0 & -58.31\\
    &&&& $\Delta t$=4s, $\Delta t_{\mathrm{BG}}$=8s&& $\Delta t$=0.25s, $\Delta t_{\mathrm{BG}}$=32s&\\

    S190924h    & Disabled &-&-&-&-&- & - \\
    
    S190930s    & No trigger & 100 \% & 3.26& HXM2, High, ch2-3 & 5.37 & HXM2, High, ch0 & 30.15\\
    &&&& $\Delta t$=0.5s, $\Delta t_{\mathrm{BG}}$=16s&& $\Delta t$=4s, $\Delta t_{\mathrm{BG}}$=8s&\\
    
    S190930t    & No trigger  & 74 \% & 3.86 & SGM, High, ch2-3 &5.07 & HXM2, Low, ch2 & -37.62\\
    &&&& $\Delta t$=4s, $\Delta t_{\mathrm{BG}}$=8s&& $\Delta t$=0.125s, $\Delta t_{\mathrm{BG}}$=32s&\\

    S191105e    & Disabled &-&-&-&-&- &- \\
    S191109d    & Disabled &-&-&-&-&- &-\\
    
    S191129u    & No trigger & 70 \% & 3.23 & HXM2, High, ch0-1 & 4.65 & HXM2, Low, ch0 & 32.93\\
    &&&& $\Delta t$=4s, $\Delta t_{\mathrm{BG}}$=8s&& $\Delta t$=0.5s, $\Delta t_{\mathrm{BG}}$=8s&\\
    
    S191204r    & No trigger & 4 \% & 4.19 & HXM1, High, ch1 & 5.19 & SGM, Low, ch0 & -5.95\\
    &&&& $\Delta t$=4s, $\Delta t_{\mathrm{BG}}$=8s&& $\Delta t$=0.125s, $\Delta t_{\mathrm{BG}}$=8s&\\
    
    S191205ah   & Disabled & - &-&-&-&-&- \\
    
    S191213g    & No trigger & 71 \% & 4.36 & HXM2, Low, ch0-3 & 5.53 & SGM, High, ch3 & 57.89\\
    &&&& $\Delta t$=4s, $\Delta t_{\mathrm{BG}}$=8s&& $\Delta t$=2s, $\Delta t_{\mathrm{BG}}$=8s&\\

    S191215w    & No trigger & 83 \% & 3.85 & HXM1, High, ch1-3 & 6.12 & HXM2, High, ch0 & 59.86\\
    &&&& $\Delta t$=4s, $\Delta t_{\mathrm{BG}}$=8s&& $\Delta t$=0.125s, $\Delta t_{\mathrm{BG}}$=8s&\\
    
    S191216ap   & No trigger & 40 \% & 2.68 & HXM1, Low, ch2 & 5.46 & HXM2, Low, ch1 & 19.65\\
    &&&& $\Delta t$=4s, $\Delta t_{\mathrm{BG}}$=8s&& $\Delta t$=0.25s, $\Delta t_{\mathrm{BG}}$=8s&\\

    S191222n    & No trigger & 60 \% & 3.02 & HXM1, Low, ch2 & 5.59 & HXM1, Low, ch2-3 & -44.56\\
    &&&& $\Delta t$=1s, $\Delta t_{\mathrm{BG}}$=8s&& $\Delta t$=2s, $\Delta t_{\mathrm{BG}}$=8s&\\

    S200105ae   & No trigger & 67 \% & 3.73 & HXM2, Low, ch0-3 & 5.85 & SGM, High, ch3 & 44.79\\
    &&&& $\Delta t$=4s, $\Delta t_{\mathrm{BG}}$=64s&& $\Delta t$=2s, $\Delta t_{\mathrm{BG}}$=64s&\\

    S200112r    & No trigger & 67 \% & 4.36 & HXM2, Low, ch1-3 &  7.16 & HXM2, High, ch0 &  -1.41\\
    &&&& $\Delta t$=4s, $\Delta t_{\mathrm{BG}}$=8s&& $\Delta t$=0.125s, $\Delta t_{\mathrm{BG}}$=64s&\\

    S200114f    & Disabled & - &-&-&-&-&- \\
    S200115j    & Disabled & - &-&-&-&-&-  \\
    
    S200128d    & No trigger & 60 \% & 3.31 & HXM2, Low, ch1 & 5.59  & HXM2, Low, ch1 & -1.29\\
    &&&& $\Delta t$=4s, $\Delta t_{\mathrm{BG}}$=8s&& $\Delta t$=0.5s, $\Delta t_{\mathrm{BG}}$=8s&\\

    S200129m    & Disabled & -&-&-&-&-&- \\
    S200208q    & Disabled & -&- &-&-&-&- \\
    
    S200213t    & No trigger & 18 \% & 4.26 & HXM2, High, ch2 & 5.27 & HXM2, High, ch3 & 20.24\\
    &&&& $\Delta t$=4s, $\Delta t_{\mathrm{BG}}$=16s&& $\Delta t$=0.125s, $\Delta t_{\mathrm{BG}}$=16s&\\
    
    S200219ac   & No trigger & 71 \% & 4.86 & HXM1, High, ch1-3 & 5.55 & HXM2, High, ch0 & 20.04\\
    &&&& $\Delta t$=4s, $\Delta t_{\mathrm{BG}}$=64s&& $\Delta t$=0.125s, $\Delta t_{\mathrm{BG}}$=32s&\\

    S200224ca   & Disabled &-&-&-&-&- & -\\
    S200225q    & Disabled &-&-&-&-&- & -\\
    
    S200302c    & No trigger & 81 \% & 3.78 & HXM1, Low, ch0 & 5.36 & HXM2, Low, ch0 & 3.56\\
    &&&& $\Delta t$=4s, $\Delta t_{\mathrm{BG}}$=8s&& $\Delta t$=1s, $\Delta t_{\mathrm{BG}}$=16s&\\
    
    S200311bg   & Disabled & - \\
    
    S200316bj   & No trigger & 90 \% & 3.08 & HXM2, Low, ch0 & 5.34 & HXM1, Low, ch0 & 12.51\\
    &&&& $\Delta t$=0.25s, $\Delta t_{\mathrm{BG}}$=8s&& $\Delta t$=0.125s, $\Delta t_{\mathrm{BG}}=$8s&\\

\hline 
\end{longtable}

\begin{table*}[t]
\caption{Conditions for SNR calculation}
\label{tab:lc_cond}
\begin{center}
\small
\begin{tabular}{lcc} \hline\hline\\[-6pt]
 & Number of conditions &  Conditions\\ \hline
detector  & 3 &  HXM1, HXM2, SGM\\
gain  & 2 &  High, Low\\
channels  & 10 &  ch0, ch1, ch2, ch3, ch0-1, ch1-2, ch2-3, ch0-2, ch1-3, ch0-3\\
$\Delta t$ & 6 & 1/8~s, 1/4~s, 1/2~s, 1~s, 2~s, 4~s\\
$\Delta t_{\mathrm{BG}}$ &  4 &8~s, 16~s, 32~s, 64~s\\
\hline 
\end{tabular}
\end{center}
\end{table*}

\newpage
\begin{longtable}{lccccccc}
\caption{Summary of CGBM/HXM1 upper limits for gravitational wave events in O3. }
\label{tab:cgbm_ul}\\
\hline\hline
Event name  & $P_{\rm HXM1}$ &$\alpha(^{\circ})$ & $\delta(^{\circ})$&  $\theta(^{\circ})$ & $\phi(^{\circ})$ & $7 \sigma$ Upper limit &  $7 \sigma$ Upper limit\\
& & &  &&& [erg cm$^{-2}$s$^{-1}$ ]  &  [erg cm$^{-2}$s$^{-1}$ ] \\ & & &  &&& (Band function)  &  (Cut-off power law) \\
\hline

\endhead    
    S190408an & 99 \% & 351.0 & 53.9 & 45 & 305 & 9.7 $\times 10^{-7}$  & 1.2 $\times 10^{-6}$  \\
    S190412m & - & - & - & - & - & - & -\\
    S190421ar & 0 \% & - & - & - & - & - & -\\
    S190425z & - & - & - &- & - & - & -\\
    S190426c & - & - & - & - & - & - & -\\
    S190503bf & - &  -  & - &- & - & - & -\\
    S190510g &2 \% & 250.3 & 15.7 & 50 & 125&  1.9 $\times 10^{-6}$ &  2.3 $\times 10^{-6}$\\
    S190512at &2 \%& 227.8 & -4.2 & 45& 155& 1.4 $\times 10^{-6}$& 1.6 $\times 10^{-6}$\\
    S190513bm &55 \%& 52.5 &  47.9 &70&10& 1.9 $\times 10^{-6}$ & 2.3 $\times 10^{-6}$\\
    S190517h &0 \%& -  & - & - & - & - & -\\
    S190519bj &3 \%& 358.2 & 51.4 & 65 & 325 & 1.8 $\times 10^{-6}$ & 2.1 $\times 10^{-6}$\\
    S190521g & - & - & - &- & - & - & -\\
    S190521r &-& - & - & - & - & - & -\\
    S190602aq &5 \%&92.8 & 8.9& 45 & 100 & 1.4 $\times 10^{-6}$  & 1.7 $\times 10^{-6}$\\
    S190630ag & - & - & - & - & - & - & -\\
    S190701ah & 0 \%& - & - & - & - & - & -\\
    S190706ai &-& - & -  & - & - & - & -\\    
    S190707q &25 \%&311.7 & 38.3& 60 & 260 & 1.8 $\times 10^{-6}$ & 2.2 $\times 10^{-6}$\\
    S190718y &9 \%& 138.1 & -31.0 & 55 & 255& $ 1.6 \times 10^{-6}$ & 1.9 $\times 10^{-6}$ \\
    S190720a & - & -  & - & - & - & - & -\\
    S190727h & 0 \%& - & - & - & - & - & -\\
    S190728q & 0 \% & - & - & - & - & - & -\\
    S190814bv & - & - & - & - & - & - & -\\
    Fermi GBM-190816 & 32 \%&183.9 & 26.6 & 40 & 115 & 1.1 $\times 10^{-6}$ & 1.3 $\times 10^{-6}$\\
    S190828j &4 \%& 342.3 & 52.9 & 50 & 330 & 9.5 $\times 10^{-7}$ & 1.2 $\times 10^{-6}$\\
    S190828l &1 \%&83.8 & 45.7 & 15 & 105 &  1.0 $\times 10^{-6}$ & 1.3 $\times 10^{-6}$\\ 
    S190901ap &4 \%&276.7& 37.2 & 70 & 0 & 1.8 $\times 10^{-6}$ & 2.2 $\times 10^{-6}$ \\
    S190910d &0 \%& - & - & - & - & - & - \\
    S190910h & 15 \%& 232.0 & -21.4 & 65 & 55 &  1.6 $\times 10^{-6}$& 2.0 $\times 10^{-6}$\\
    S190915ak &0 \%& - & - & - & - & - & - \\
    S190923y & 14 \% & 113.6 & 32.3 & 65 & 355 & 1.6 $\times 10^{-6}$ &  1.9 $\times 10^{-6}$\\
    S190924h & - & - & - & - & - & - & -\\
    S190930s & 11 \%& 328.2 & 3.6 & 55 & 130 & 1.6 $\times 10^{-6}$ & 1.9 $\times 10^{-6}$\\
    S190930t  & 11 \%&144.1 & 37.2 & 70 & 15 & 2.0 $\times 10^{-6}$ & 2.4 $\times 10^{-6}$\\
    S191105e & - & -  &  - & - & - & - & - \\ 
    S191109d & - & - & - & - & - & - & - \\
    S191129u &0 \%& - & - & - & - & - & - \\
    S191204r & 2 \%& 218.1 & 66.0 & 45 & 345 & 1.1 $\times 10^{-6}$ & 1.3 $\times 10^{-6}$\\
    S191205ah &-& - & - & - & - & - &  -  \\ 
    S191213g & 11 \%& 87.4 & -8.6 & 65 & 310 & 1.7 $\times 10^{-6}$ & 2.0 $\times 10^{-6}$ \\
    S191215w & 0 \%& - & - & - & - & - & -\\
    S191216ap &0 \%& - & - & - & - & - & - \\
    S191222n & 5 \%& 37.3 & 20.6 & 70 & 340 & 1.9 $\times 10^{-6}$ & 2.3 $\times 10^{-6}$\\
    S200105ae & 47 \%& 53.8 & -18.8 & 10 & 295 & 9.1 $\times 10^{-7}$ & 1.1 $\times 10^{-6}$\\
    S200112r & 6 \%& 101.2 & 23.2 & 20 & 185& 9.7 $\times 10^{-7}$ & 1.2 $\times 10^{-6}$ \\
    S200114f &-& - &  - & - & - & - & -\\
    S200115j &-& - & - &- & - & - & -\\
    S200128d & 22 \%& 64.7 & 36.2 & 55 & 105& 1.5 $\times 10^{-6}$ & 1.8 $\times 10^{-6}$\\
    S200129m &-& - & - & - & - & - & -\\
    S200208q &-& - & -& - & - & - & -\\
    S200213t & 4 \% & 180.1 & -37.9 & 60 & 280 & 1.5 $\times 10^{-6}$ & 1.8 $\times 10^{-6}$ \\
    S200219ac & 18 \% & 185.1 & 56.5 & 60 & 25 & 1.6 $\times 10^{-6}$ & 1.9 $\times 10^{-6}$\\
    S200224ca &-& - & - & - & - & - & -\\
    S200225q & - & - & - & - & - & - & -\\
    S200302c & 22 \% &36.7 & 69.1 & 55 & 340 &1.5 $\times 10^{-6}$ & 1.8 $\times 10^{-6}$ \\
    S200311bg &-& - & - & - & - & - & -\\
    S200316bj & 13 \% &90.0 & 46.5  & 35 & 45 & 1.0 $\times 10^{-6}$ & 1.3 $\times 10^{-6}$\\
\hline 
\end{longtable}

\newpage
\begin{longtable}{lccccccc}
\caption{Summary of CGBM/HXM2 upper limits for gravitational wave events in O3. }
\label{tab:cgbm_ul}\\
\hline\hline
Event name  & $P_{\rm HXM2}$ &$\alpha(^{\circ})$ & $\delta(^{\circ})$&  $\theta(^{\circ})$ & $\phi(^{\circ})$ & $7 \sigma$ Upper limit &  $7 \sigma$ Upper limit\\
& & &  &&& [erg cm$^{-2}$s$^{-1}$ ]  &  [erg cm$^{-2}$s$^{-1}$ ] \\
& & &  &&& (Band function)  &  (Cut-off power law) \\ \hline

\endhead    
    S190408an & 99 \% & 351.0 & 53.9 & 45 & 305 & 1.2 $\times 10^{-6}$  & 1.5 $\times 10^{-6}$  \\
    S190412m & - & - & - & - & - & - & -\\
    S190421ar & 0 \% & - & - & - & - & - & -\\
    S190425z & - & - & - &- & - & - & -\\
    S190426c & - & - & - & - & - & - & -\\
    S190503bf & - &  -  & - &- & - & - & -\\
    S190510g &2 \% & 250.3 & 15.7 & 50 & 125&  1.7 $\times 10^{-6}$ &  2.0 $\times 10^{-6}$\\
    S190512at &1 \%& 226.8 & -5.1 & 45 & 155 & 1.6 $\times 10^{-6}$& 1.9 $\times 10^{-6}$\\
    S190513bm &55 \%& 52.5 &  47.9 & 70 & 10 & 1.7 $\times 10^{-6}$ & 2.0 $\times 10^{-6}$\\
    S190517h &0 \%& -  & - & - & - & - & -\\
    S190519bj &3 \%& 359.9 & 53.7 & 65 & 330 & 1.8 $\times 10^{-6}$ & 2.2 $\times 10^{-6}$\\
    S190521g & - & - & - &- & - & - & -\\
    S190521r &-& - & - & - & - & - & -\\
    S190602aq &5 \%&92.8 & 8.9& 45 & 100 & 1.5 $\times 10^{-6}$  & 1.8 $\times 10^{-6}$\\
    S190630ag & - & - & - & - & - & - & -\\
    S190701ah & 0 \%& - & - & - & - & - & -\\
    S190706ai &-& - & -  & - & - & - & -\\
    S190707q &26 \%&311.7 & 38.3& 60 & 260 & 2.0 $\times 10^{-6}$ & 2.4 $\times 10^{-6}$\\
    S190718y &9 \%& 138.1 & -31.0 & 55 & 255 & 1.5 $\times 10^{-6}$ & 1.8 $\times 10^{-6}$ \\
    S190720a & - & -  & - & - & - & - & -\\
    S190727h & 0 \%& - & - & - & - & - & -\\
    S190728q & 0 \% & - & - & - & - & - & -\\
    S190814bv & - & - & - & - & - & - & -\\
    Fermi GBM-190816 & 34 \%&183.9 & 26.6 & 40 & 115 & 1.0 $\times 10^{-6}$ & 1.2 $\times 10^{-6}$\\
    S190828j &4 \%& 339.6 & 51.2 & 50 & 335 & 1.1 $\times 10^{-6}$ & 1.3 $\times 10^{-6}$\\
    S190828l &1 \%&83.8 & 45.7 & 15 & 105 &  1.1 $\times 10^{-6}$ & 1.4 $\times 10^{-6}$\\ 
    S190901ap &5 \%&277.4& 33.5 & 70 & 5& 1.7 $\times 10^{-6}$ & 2.0 $\times 10^{-6}$ \\
    S190910d &0 \%& - & - & - & - & - & - \\
    S190910h & 16 \%& 232.0 & -24.0 & 65 & 60 &  1.8 $\times 10^{-6}$& 2.2 $\times 10^{-6}$\\
    S190915ak &0 \%& - & - & - & - & - & - \\
    S190923y & 16 \% & 111.6 & 31.0 & 65 & 355 & 1.6 $\times 10^{-6}$ &  1.9 $\times 10^{-6}$\\
    S190924h & - & - & - & - & - & - & -\\
    S190930s & 12 \%& 328.2 & 0.6 & 55 & 135 & 1.6 $\times 10^{-6}$ & 2.0 $\times 10^{-6}$\\
    S190930t  & 12 \% & 144.8 & 36.4 & 70 & 15 & 1.7 $\times 10^{-6}$ & 2.0 $\times 10^{-6}$\\    
    S191105e & - & -  &  - & - & - & - & - \\ 
    S191109d & - & - & - & - & - & - & - \\
    S191129u &0 \%& - & - & - & - & - & - \\
    S191204r & 3 \%& 219.3 & 65.5 & 45 & 345 & 1.1 $\times 10^{-6}$ & 1.3 $\times 10^{-6}$\\
    S191205ah &-& - & - & - & - & - & - \\ 
    S191213g & 11 \%& 87.4 & -8.6 & 65 & 310 & 1.7 $\times 10^{-6}$ & 2.0 $\times 10^{-6}$ \\
    S191215w & 0 \%& - & - & - & - & - & -\\
    S191216ap &0 \%& - & - & - & - & - & - \\
    S191222n & 5 \%& 37.3 & 20.6 & 70 & 340 & 1.8 $\times 10^{-6}$ & 2.1$\times 10^{-6}$\\
    S200105ae & 52 \%& 53.8 & -18.8 & 10 & 295 & 1.0 $\times 10^{-6}$ & 1.2 $\times 10^{-6}$\\
    S200112r & 8 \%& 109.7 & 15.6 & 35 & 190 & 1.2 $\times 10^{-6}$ & 1.4 $\times 10^{-6}$\\
    S200114f &-& - &  - & - & - & - & -\\
    S200115j &-& - & - &- & - & - & -\\
    S200128d & 23 \%& 62.6 & 34.0 & 55 & 105 & 1.3 $\times 10^{-6}$ & 1.6 $\times 10^{-7}$\\
    S200129m &-& - & - & - & - & - & -\\
    S200208q &-& - & -& - & - & - & -\\
    S200213t & 5 \% & 180.1 & -37.9 & 60 & 280 & 1.5 $\times 10^{-6}$ & 1.8 $\times 10^{-6}$ \\
    S200219ac & 19 \% & 185.1 & 56.5 & 60 & 25 & 1.5 $\times 10^{-6}$ & 1.8 $\times 10^{-6}$\\
    S200224ca &-& - & - & - & - & - & -\\
    S200225q & - & - & - & - & - & - & -\\    
    S200302c & 22 \% &36.7 & 69.1 & 55 & 340 &1.5 $\times 10^{-6}$ & 1.8 $\times 10^{-6}$ \\
    S200311bg &-& - & - & - & - & - & -\\
    S200316bj & 14 \% &85.3 & 49.9  & 40 & 35 & 1.1 $\times 10^{-6}$ & 1.4 $\times 10^{-6}$\\
\hline 
\end{longtable}

\newpage
\begin{longtable}{lccccccc}
\caption{Summary of CGBM/SGM upper limits for gravitational wave events in O3. }
\label{tab:cgbm_ul}\\
\hline\hline
Event name  & $P_{\rm SGM}$ &$\alpha(^{\circ})$ & $\delta(^{\circ})$&  $\theta(^{\circ})$ & $\phi(^{\circ})$ & $7 \sigma$ Upper limit &  $7 \sigma$ Upper limit\\
& & &  &&& [erg cm$^{-2}$s$^{-1}$ ]  &  [erg cm$^{-2}$s$^{-1}$ ] \\
& & &  &&& (Band function)  &  (Cut-off power law) \\
 \hline

\endhead    
    S190408an & 100 \% & 351.0 & 53.9 & 45 & 305 & 8.2 $\times 10^{-7}$  & 9.3 $\times 10^{-7}$  \\
    S190412m & - & - & - & - & - & - & -\\
    S190421ar & 0 \% & - & - & - & - & - & -\\
    S190425z & - & - & - &- & - & - & -\\
    S190426c & - & - & - & - & - & - & -\\
    S190503bf & - &  -  & - &- & - & - & -\\
    S190510g &4 \% & 205.3 & 0.6 & 90 & 100 &  1.2 $\times 10^{-6}$ &  1.3 $\times 10^{-6}$\\
    S190512at &3 \%& 228.2 & -5.1 & 45 & 155 & 8.7 $\times 10^{-7}$& 1.0 $\times 10^{-6}$\\
    S190513bm &76 \%& 52.6 &  47.8 & 70 & 5 & 1.0 $\times 10^{-6}$ & 1.2 $\times 10^{-6}$\\
    S190517h &0 \%& -  & - & - & - & - & -\\
    S190519bj &8 \%& 355.3 & 48.2 & 65 & 320 & 1.1 $\times 10^{-6}$ & 1.2 $\times 10^{-6}$\\
    S190521g & - & - & - &- & - & - & -\\
    S190521r &-& - & - & - & - & - & -\\
    S190602aq &21 \%&72.6 & -10.7& 75 & 95 & 1.1 $\times 10^{-6}$  & 1.3 $\times 10^{-6}$\\
    S190630ag & - & - & - & - & - & - & -\\
    S190701ah & 0 \%& - & - & - & - & - & -\\
    S190706ai &-& - & -  & - & - & - & -\\
    S190707q &48 \%&175.4 &  -48.3& 90 & 85 & 9.5 $\times 10^{-7}$ & 1.1 $\times 10^{-6}$\\
    S190718y &14 \%& 137.1& -29.4 & 55 & 250 & 7.8 $\times 10^{-7}$ & 8.9 $\times 10^{-7}$ \\
    S190720a & - & -  & - & - & - & - & -\\
    S190727h & 8 \%&353.7 & 52.3& 85 & 300 & 1.3 $\times 10^{-6}$ & 1.5 $\times 10^{-6}$\\
    S190728q & 0 \% & - & - & - & - & - & -\\
    S190814bv & - & - & - & - & - & - & -\\
    Fermi GBM-190816 & 36 \%&183.9 & 26.6 & 40 & 115 & 8.7 $\times 10^{-7}$ & 9.9 $\times 10^{-7}$\\
    S190828j &4 \%& 344.0 & 54.3 & 50 & 330 & 8.5 $\times 10^{-7}$ & 9.7 $\times 10^{-7}$\\
    S190828l &3 \%&348.0 & 32.0 & 80 & 55 &  1.2 $\times 10^{-6}$ & 1.4 $\times 10^{-6}$\\ 
    S190901ap &20 \%&257.7 & 20.1& 90 & 15 & 1.3 $\times 10^{-6}$ & 1.5 $\times 10^{-6}$ \\    
    S190910d &1 \%& 340.7 & 54.9 & 85 & 75 & 9.5 $\times 10^{-7}$ & 1.1 $\times 10^{-6}$\\
    S190910h & 26 \%& 220.1 & -22.0 & 75 & 55 &  9.8 $\times 10^{-7}$& 1.1 $\times 10^{-6}$\\
    S190915ak &0 \%& - & - & - & - & - & -\\
    S190923y & 26 \% &103.9 & 24.6 & 55 & 350 & 8.7 $\times 10^{-7}$ &  9.9 $\times 10^{-7}$\\
    S190924h & - & - & - & - & - & - & -\\
    S190930s & 37 \%& 322.5 & 48.5 & 70 & 85 & 9.5 $\times 10^{-7}$ & 1.1 $\times 10^{-6}$\\
    S190930t  & 22 \%&136.4 & 30.7 &  80 & 15 & 1.5 $\times 10^{-6}$ & 1.7 $\times 10^{-6}$\\
    S191105e & - & -  &  - & - & - & - & - \\ 
    S191109d & - & - & - & - & - & - & - \\
    S191129u &4 \%& 201.8 &  41.7 & 85 & 20 & 1.4 $\times 10^{-6}$ & 1.6 $\times 10^{-6}$\\
    S191204r & 2 \%& 200.7 &  38.3 & 55 & 20 & 9.0 $\times 10^{-7}$ & 1.0 $\times 10^{-6}$\\    
    S191205ah &-& - & - & - & - & - & -\\ 
    S191213g & 35 \%&106.0 & 0.1 & 85 & 320 & 1.4 $\times 10^{-6}$ & 1.5 $\times 10^{-6}$ \\
    S191215w & 55 \%& 325.8 &  20.1& 85 & 265 & 1.2 $\times 10^{-6}$ & 1.4 $\times 10^{-6}$\\
    S191216ap &4 \%& 299.0 & 57.1& 90 & 275 & 3.2 $\times 10^{-6}$ & 3.3 $\times 10^{-6}$ \\
    S191222n & 14 \%& 54.3 & 39.1 & 85 & 355 & 1.2 $\times 10^{-6}$ & 1.4 $\times 10^{-6}$\\
    S200105ae & 50 \%& 53.8 & -18.8 & 10 & 295 & 9.3 $\times 10^{-7}$ & 1.1 $\times 10^{-6}$\\
    S200112r & 16 \%& 260.1 &  50.8 &90 & 315 & 9.6 $\times 10^{-7}$ & 1.1 $\times 10^{-6}$\\
    S200114f &-& - &  - & - & - & - & -\\
    S200115j &-& - & - &- & - & - & -\\
    S200128d & 25 \%& 64.3 & 35.9 & 55 & 105 & 8.1 $\times 10^{-7}$ & 9.2 $\times 10^{-7}$\\
    S200129m &-& - & - & - & - & - & -\\
    S200208q &-& - & -& - & - & - & -\\
    S200213t & 7 \% & 180.4 & -37.6 & 60 & 280 & 7.4 $\times 10^{-7}$ & 8.4 $\times 10^{-7}$ \\
    S200219ac & 20 \% & 185.1 & 56.5 &60 & 25 & 1.1 $\times 10^{-6}$ & 1.2 $\times 10^{-6}$\\
    S200224ca &-& - & - & - & - & - & -\\
    S200225q & - & - & - & - & - & - & -\\
    S200302c & 35 \% &36.7 & 69.1 & 55 & 340 &9.4 $\times 10^{-7}$ & 1.1 $\times 10^{-6}$ \\
    S200311bg &-& - & - & - & - & - & -\\
    S200316bj & 14 \% &90.0 & 46.5  & 35 & 45 & 8.3 $\times 10^{-7}$ & 9.5 $\times 10^{-7}$\\
\hline 
\end{longtable}

\newpage
\begin{longtable}{lccccc}
\caption{Summary of CAL observations for gravitational wave events in O3. }
\label{tab:cal_obs}\\
\hline\hline
    Event name  & $\alpha(^{\circ})$ & $\delta(^{\circ})$ &  Coverage & Run mode & 90\% Upper limit \\
    & &  & & & [erg cm$^{-2}$s$^{-1}$ ]  \\ \hline
\endhead
    
    S190408an   &  352.9 & 8.4  & 95 \% & LEG & 3.0 $\times 10^{-7}$\\
    S190412m    & - & - & - & - & - \\
    S190421ar   & 326.6 & 42.3 & 0 \% & -  & - \\
    S190425z    & 131.4 & -43.7 & 10 \% &HE & 8.5 $\times 10^{-5}$\\
    S190426c    & 183.1 & -50.9 & 10 \%& HE & 9.2 $\times 10^{-6}$\\
    S190503bf   & 169.1 & -45.5 & 25 \% & HE & 7.1 $\times 10^{-5}$\\
    S190510g    & 295.8 & 50.8 & 0 \%& - & -\\    
    S190512at   & 214.9 & 37.8 & 0 \% & - & - \\
    S190513bm   & 348.0 & 4.3 & 15 \%& LEG & 4.5 $\times 10^{-5}$\\    
    S190517h    & 125.9 & -31.5 & 0 \% & - & - \\
    S190519bj   & 243.4 & 51.1 & 0 \% & - & - \\
    S190521g    & 205.8 & 49.3 & 30 \% & HE& 7.4 $\times 10^{-7}$\\    
    S190521r     & 225.4 & 51.4 & 0 \% & - & - \\
    S190602aq   & 127.3 & 45.7 & 0 \%& - & - \\
    S190630ag   & 84.0 & 31.5 & 0 \%& - & - \\
    S190701ah   & 286.9 & -1.6 & 0 \% & - & - \\
    S190706ai   & 210.4 & -45.4 & 0 \%& - & - \\
    S190707q    & 262.4 & 2.2 & 25 \% & LEG & 3.8 $\times 10^{-6}$ \\  
    S190718y    & 195.8 & -11.0 & 10 \% & LEG & 1.2 $\times 10^{-5}$ \\
    S190720a    & 62.9 & -40.5 & 0 \% & - & - \\
    S190727h    & 201.2 & 38.3 & 0 \% & - & - \\
    S190728q    & 184.9 & 30.3 & 0 \% & - & - \\
    S190814bv   & 182.7& 49.2 & 0 \% & - & - \\
    Fermi GBM-190816 & 227.4 & 14.7 & 25 \% & HE & 2.8 $\times 10^{-5}$ \\
    S190828j    & 13.9 & 12.7 & 0 \% &- & - \\
    S190828l    & 107.1 & 51.0 & 0 \% & - & - \\ 
    S190901ap   & 353.8 & 16.7 & 5 \% & LEG & 2.8 $\times 10^{-5}$ \\
    S190910d    & 100.9 & 22.9 &0 \% & - & - \\
    S190910h    & 294.8 & -5.4 & 10 \% & LEG & 5.3 $\times 10^{-7}$ \\
    S190915ak   & 99.8 & -11.1 & 0 \% & - & - \\
    S190923y    & 55.3 & -2.6 & 0 \% & - & - \\    
    S190924h    & 273.5 & 40.2 & 0 \% & - & - \\    
    S190930s    & 20.8 & -3.4 & 5 \% & HE & 4.5 $\times 10^{-5}$ \\
    S190930t    & 235.5 & 36.3 & 0 \% & - & - \\
    S191105e    & 223.0 & -27.4 & 0 \% & - & - \\    
    S191109d    & 349.8 & -16.6 & 0 \% & - & - \\
    S191129u    & 356.8 & 50.7 & 0\% & - & - \\
    S191204r    & 269.2 & 34.3 & 0 \% & - & - \\
    S191205ah   & 80.2 & -32.8 & 0\% & - & - \\
    S191213g    & 20.4 & -9.3 & 5 \% & LEG & 1.5 $\times 10^{-5}$  \\
    S191215w    & 222.3 & 40.3 & 0 \% & - & - \\
    S191216ap   & 186.8 & 13.9 & 0 \% & - & - \\
    S191222n    & 330.3 & -2.1 & 0 \% & - & - \\
    S200105ae   & 50.6 & -30.6 & 45 \% & HE & 3.1 $\times 10^{-5}$ \\  
    S200112r    & 84.6 & 40.0 & 5 \% & HE & 1.1 $\times 10^{-6}$ \\
    S200114f    & 111.1 & 50.7 & 85 \% & HE &  1.2 $\times 10^{-5}$\\
    S200115j    & 84.4 & 45.9 & 15 \% & HE & 8.5 $\times 10^{-5}$\\
    S200128d    & 126.1 & 23.4 & 5 \%  & HE & 4.5 $\times 10^{-6}$\\
    S200129m    & 288.7 & -34.3 & 5 \% & HE & 4.8 $\times 10^{-4}$\\
    S200208q    & 224.1 & -41.8 & 0 \% & - & - \\
    S200213t    & 101.4 & -36.1 & 0 \% & - & - \\
    S200219ac   & 298.4 & 51.6 & 0 \% & - & - \\
    S200224ca   & 167.5 & -24.8 & 95 \% & HE & 9.0  $\times 10^{-7}$\\    
    S200225q    & 157.6 & -32.3 & 0 \% & - & - \\
    S200302c    & 245.6 & 52.0 & 0 \% & - & - \\
    S200311bg   & 191.3 & 51.5 & 0 \% & - & - \\
    S200316bj   & 144.7 & 47.5 & 0 \% & - & - \\
\hline 
\end{longtable}

\bibliography{gw}{}
\bibliographystyle{aasjournal}



\end{document}